\documentclass[conference]{IEEEtran}

\usepackage[utf8]{inputenc} 
\usepackage[T1]{fontenc}    
\usepackage{hyperref}       
\usepackage{url}            
\usepackage{booktabs}       
\usepackage{amsfonts}       
\usepackage{nicefrac}       
\usepackage{doi}
\usepackage{orcidlink}
\usepackage{marvosym}
\usepackage{authblk}
\usepackage{graphicx}
\usepackage{bm}
\usepackage{amssymb}
\usepackage{gensymb}
\usepackage{tabularx}
\usepackage{caption}
\usepackage{subcaption}
\usepackage{algorithm}
\usepackage{algpseudocode}
\usepackage{amsmath}
\usepackage{multirow}
\usepackage{colortbl}
\usepackage{etoolbox}
\newcommand{\insertfig}{
  \begin{center}
    \setcounter{figure}{0}
    \captionsetup{belowskip=0pt}
    \includegraphics[width=\textwidth]{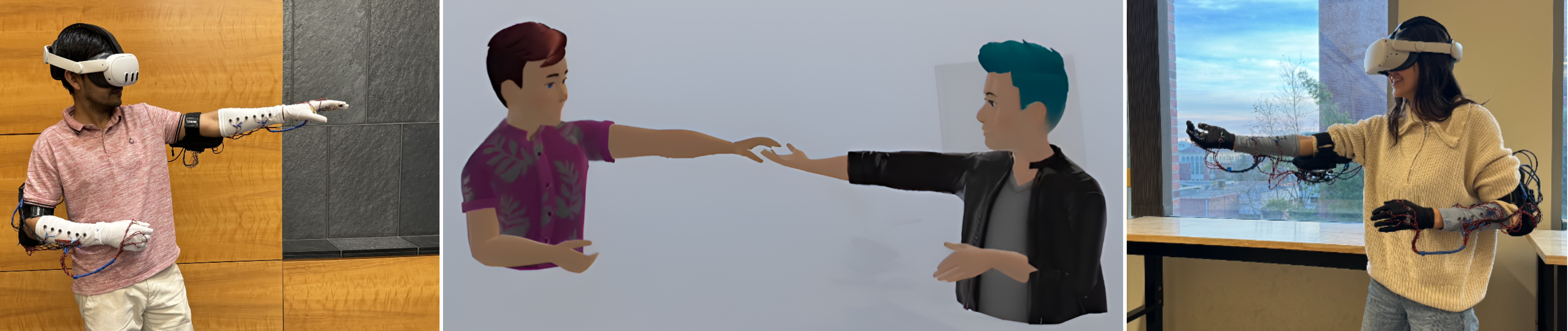}
    \captionof{figure}{Real-time haptic interaction between two physically distant users within the same VR environment}
  \end{center}
}

\makeatletter
\apptocmd{\@maketitle}{\centering\insertfig}{}{}
\makeatother

\begin{document}
\title{Virtual Encounters of the Haptic Kind: Towards a Multi-User VR System for Real-Time Social Touch}
\author[1]{Premankur Banerjee\orcidlink{0000-0002-0865-3634} \thanks{\href{mailto:premankur.banerjee@usc.edu}{premankur.banerjee@usc.edu}}}
\author[1]{Jiaxuan Wang\orcidlink{}}
\author[2]{Lauren Tomita\orcidlink{0009-0009-9245-2007}}
\author[2]{Mia P Montiel\orcidlink{0009-0000-7972-1166}}
\author[1]{Heather Culbertson\orcidlink{0000-0002-9187-2706}}

\affil[1]{Thomas Lord Department of Computer Science}
\affil[2]{Alfred E. Mann Department of Biomedical Engineering}
\affil[1,2]{Viterbi School of Engineering, University of Southern California, Los Angeles, CA 90089, USA}

\date{}


\hypersetup{
pdftitle={Virtual Encounters of the Haptic Kind: Towards a Multi-User VR System for Real-Time Social Touch},
pdfauthor={Premankur Banerjee, Jiaxuan Wang, Lauren Tomita, Mia P Montiel, Heather Culbertson},
pdfkeywords={Virtual Reality, Motion Simulation, Acceleration Perception, Surfing},
}

\maketitle

\begin{abstract}
    Physical touch, a fundamental aspect of human social interaction, remains largely absent in real-time virtual communication. We present a haptic-enabled multi-user Virtual Reality (VR) system that facilitates real-time, bi-directional social touch communication among physically distant users.
We developed wearable gloves and forearm sleeves, embedded with 26 vibrotactile actuators for each hand and arm, actuated via a WiFi-based communication system. The system enables VR-transmitted data to be universally interpreted by haptic devices, allowing feedback rendering based on their capabilities. 
Users can perform and receive social touch gestures such as stroke, pat, poke, and squeeze, with other users within a shared virtual space or interact with other virtual objects, and they receive vibrotactile feedback.
Through a two-part user study involving six pairs of participants, we investigate the impact of gesture speed, haptic feedback modality, and user roles, during real-time haptic communication in VR, on affective and sensory experiences, as well as evaluate the overall system usability.
Our findings highlight key design considerations that significantly improve affective experiences, presence, embodiment, pleasantness, and naturalness, to foster more immersive and expressive mediated social touch experiences in VR.
\end{abstract}



\section{Motivation}

The COVID-19 pandemic and the resulting need for social distancing~\cite{DH3, DH4, DH5, DH6} significantly accelerated the adoption of social Virtual Reality (VR) as a medium for virtual communication. Applications such as virtual meetings and negotiations~\cite{DH7, DH8, DH9, qiu2023vigather} have demonstrated the potential of VR to improve convenience and productivity in human interaction. However, these virtual environments often struggle to effectively convey social intentions~\cite{DH10, DH11}, particularly non-verbal behaviors, which are estimated to have five times the impact of verbal communication on expressing emotional connections~\cite{DH12}. This limitation is particularly pronounced in the absence of physical touch -- a critical element of human interaction that fosters social bonding and connection~\cite{DH13, DH14}.

Physical touch is an evolutionarily fundamental mode of interaction~\cite{ST8}, fulfilling an innate human need for social contact. Mediated touch, which uses technology to replicate touch sensations in remote settings, has gained considerable attention in recent years~\cite{ST27,wei2023mediated}. The growing interest has been further fueled by the emergence of the Metaverse, a conceptual framework for an integrated, immersive ecosystem where the boundaries between the virtual and real worlds are seamless~\cite{DH43, DH44}. The Metaverse seeks to improve the psychological and emotional engagement of users, highlighting the importance of replicating the social presence and emotional connection in real-world and virtual environments.
The increasing need for remote social interaction and the evolution of the Metaverse underscore the need to integrate mediated touch into virtual communication systems. Such technologies must not only provide a sense of presence, but also enable the authentic transmission of social intentions and emotions. However, substantial challenges remain in accurately and effectively transmitting genuine emotions and intentions through mediated touch platforms~\cite{ST27}.

This paper presents a multiplayer VR application supporting up to 16 geographically distributed users, enabling real-time social interaction with their avatars as well as collaboration with other virtual objects in a shared virtual space. Haptic feedback corresponding to the interactions is rendered through 26 vibrotactile actuators or Eccentric Rotating Mass (ERM) motors distributed across each hand and forearms. The application can be used on Meta VR headsets and uses a WiFi-based communication system. It is designed to ensure that the data transmitted from VR can be universally interpreted by various haptic devices, to render feedback in accordance with their specific rendering capabilities.

\subsection{Affective Touch Technology}

Efforts to develop devices capable of rendering affective haptic stimuli have explored various mechanisms and body sites, with several approaches targeting specific use cases like social touch communication, affective messaging, enhancing multimedia experiences, among others~\cite{ALT17, ALT18, ALT19, ALT20, ALT21, ALT22, ALT27}. Many devices focus on the forearm due to the ease of mounting actuators and the appropriateness of this location for social touch~\cite{suvilehto2015topography}. The modalities leveraged by these systems, including vibration~\cite{ALT16, ALT28, ALT29, ALT30, kirchner2023phantom}, slow or static pressure cues~\cite{ALT31}, force~\cite{wang2012keep}, or temperature~\cite{he2024affective}, determine how realistic and/or expressive the affective touch is. Their actuation techniques, such as linear actuators~\cite{ALT32}, pneumatic actuators~\cite{ALT33}, voice coil actuators~\cite{ALT7, ALT34}, and mid-air~\cite{he2024affective}, determine scalability and wearability of the device. Vibrational illusions~\cite{ALT45, ALT44, tactilebrush}, in particular, offer versatility, enabling arbitrary continuous movement patterns, including in two-dimensions~\cite{ALT30, kirchner2023phantom}, making them well-suited for reproducing social touch.
Efforts to enhance mediated touch have focused on enriching the range of emotions and sensations conveyed~\cite{ALT18}. However, attempts to create more authentic reproductions of social touch often encounter trade-offs, including limitations in versatility, latency, and bulkiness compared to vibratory approaches.

A recent study~\cite{kirchner2023phantom} introduced a novel vibrotactile armband system that leverages phantom illusion and parametric design to render affective touch patterns on the forearm, which supports live rendering and automatic generation of touch patterns. However, the system would require a sensor armband to capture and recognize the touch patterns before the touch parameters are transmitted and rendered on the receiver armband. This would still not enable seamless, naturalistic feedback for the toucher in virtual environments, when they perform the gestures directly with their hands on the touchee. Sensors and communication play a critical role in mediated social touch, as they determine the fidelity of feedback and its emotional interpretation. High-quality sensors enable precise capture of input gestures like squeezes or strokes~\cite{rantala2013touch}, while effective communication systems ensure accurate transmission of these gestures to the receiver, adhering to the intended affective intent~\cite{bailenson2007virtual}. Previous approaches to two-dimensional touch pattern reproduction have often been based on recordings of real touches~\cite{ALT7, ALT28, ALT35}. Touchers aim to convey specific emotions such as delight, anger, or relaxation through tailored gestures like squeezing or stroking~\cite{huisman2013towards,smith2007communicating}, and the type of gesture often correlates with the intended valence and arousal of the emotion~\cite{rantala2013touch}.

To further enhance the authenticity of mediated social touch, it is important to integrate visual information, since body language is a vital component of non-verbal communication~\cite{bailenson2005digital}. Incorporating visual cues, such as gestures, postures, or facial expressions, together with haptic feedback, could create richer, more holistic representations of social interactions~\cite{maloney2020talking}. Visual information significantly influences the perception of affect and intention during communication~\cite{zamuner2013role, blom2021perceiving}. Thus, combining visual and haptic modalities offers a promising avenue to bridge the gap between mediated and naturalistic social interactions, enabling a more immersive and emotionally expressive experience.

\subsection{Social Touch in Virtual Reality}

Modern VR technology has demonstrated the ability to induce a strong illusion of virtual body ownership, often referred to as a sense of embodiment~\cite{ST28, ST34, ST40, ST43, ST49, ST50}. This phenomenon allows users to perceive interactions with their virtual body as if they were occurring with their physical body. For instance, observing one's virtual body being touched or interacting with another avatar can elicit reactions akin to those experienced during physical touch in the real world~\cite{jacucci2024haptics,genay2021being}. Studies further indicate that technologically mediated social touch can evoke physiological, emotional, and behavioral responses comparable to real-world touch~\cite{ST18, ST27, ST60}.

Research into social touch in VR categorizes findings based on the type of interaction partners (human-human vs. human-agent) and the direction of the touch (participant-initiated vs. participant-received).
Rendering touch using artificial hands were shown to enhance human-likeness in avatars~\cite{ST26} (human-agent, receiving touch), while force-feedback devices revealed variations in touch force strength and duration based on a virtual agent's characteristics, touch location, and participant factors like sex and anti-fat attitudes~\cite{ST2, ST59} (human-agent, initiating touch). These findings align with results from face-to-face studies in this field~\cite{ST15, ST46}, suggesting similar underlying behavioral dynamics.
The perceived appropriateness and erogeneity of visual-only virtual touch on different zones of an embodied avatar (human-agent, receiving touch)~\cite{ST14} were influenced by factors like touch location, sex of the touching avatar and the participant, and the participant’s sexual orientation. 
The role of touch in economic decision-making (human-agent, receiving and initiating touch) reported no effect of touch on compliance behavior~\cite{ST55}, in contrast to previous findings in VR settings involving human-agent interactions and receiving touch~\cite{ST21, ST65}.
The perception of virtual touch supported by tactile feedback (human-agent, receiving) was found to be modulated by the facial expressions of virtual agents as well as individual differences related to the participants’ sex~\cite{ST20}. Haptic feedback via ultrasonic arrays and silicone hand was also found to enhance the affective perception of the toucher (human-agent, initiating touch) during social touch~\cite{toucherFeedback}.

For human-human initiating and receiving interactions, there has been significant research into replicating remote handshakes in virtual environments using a robot to provide the haptic feedback
~\cite{DH32, DH33, DH34}. These systems facilitate remote communication~\cite{DH35, DH36, DH37}, especially by incorporating tactile modalities to achieve multimodality~\cite{DH38}. Haptic feedback devices have also been developed to provide remote handshake sensations~\cite{tong2024distant, DH39, DH40, DH42}.

Emotional responses to social touch in a two-user VR scenario (human-human, receiving and initiating) were investigated in~\cite{sykownik2020experience}, highlighting how factors like intimacy, touch direction, and participant sex influence these reactions. Although emotional responses mirrored those of physical touch, the study relied on minimal haptic feedback, using a simple vibration from a Vive VR controller for the person initiating touch (toucher) only. In~\cite{ST7}, vibrotactile feedback provided to a user's shoulder (touchee only) in multi-user VR (human-human, receiving) did not increase willingness to engage in embarrassing social situations; however, it enhanced the perceived realism of touch, emphasizing the importance of tactile cues in virtual interactions. A more recent study~\cite{peng2024impact} (human-human, receiving and initiating) explored the effects of vibrotactile feedback on Autonomous Sensory Meridian Response (ASMR) experiences in a two-person VR environment. The findings showed that vibrotactile feedback enhances relaxation, comfort, and enjoyment while increasing avatar embodiment and immersion. In this case, only the touchee (viewer) received tactile feedback through a bHaptics vest, arm, and gloves, while the toucher (ASMRtist) did not.

This review highlights that previous studies have rarely examined emotional responses in direct human-to-human interactions within a receiving-and-initiating setting, where real-time haptic feedback is provided simultaneously to both the toucher and the touchee in VR. Most existing research has focused on aspects such as the haptic experience of touch, touching behavior, and behavioral responses.
To the best of our knowledge, no VR application utilizing devices from the aforementioned works has been developed to enable live, real-time multi-user human-to-human mediated touch. Specifically, systems capable of providing multi-point haptic feedback for both the toucher and the touchee, allowing them to convey or receive emotions or perform gestures such as stroking, patting, or poking in VR, have not yet been demonstrated or systematically studied.

\subsection{Research Questions}

A two-part experimental study was conducted to address three primary research questions: 
\textbf{RQ1}: \textit{How effectively can users convey social touch gestures remotely through the VR system with and without tactile feedback?}
\textbf{RQ2}: \textit{What are the affective and sensory experiences associated with remote social touch gestures, both from the perspective of the initiator (toucher) and the recipient (touchee) in real-time VR interactions?}
\textbf{RQ3}: \textit{How does the VR system perform in terms of usability, user embodiment, sense of presence, and cybersickness?}


\section{System Design}

\subsection{Multiplayer VR Environment}
We developed our multi-user VR system using Unity 3D, targeting the Meta Quest headsets for rendering. Our system supports both controller inputs and hand tracking, leveraging Meta's hand tracking technology. For avatar representation, we use Meta's Avatar SDK, which allows users to select and customize their avatars. We integrated Photon Fusion, a high-performance networking engine that supports seamless synchronization and networked physics, for real-time multiplayer interactions. Photon Fusion enables multiplayer capabilities and provides built-in voice chat functionality, so users can communicate verbally within the virtual environment.

Multiple users can join our virtual space from different physical or geographical locations, provided that they have the application on their headset and a stable internet connection. The VR application we developed is designed to be a standalone application on the Meta headsets and can also be run in Quest Link or Air Link modes. The VR environment is designed as a simple room that contains a table with various interactable objects that all players can manipulate collaboratively or individually. The main aim of the VR application is to facilitate multi-user human-human social touch. Users are free to navigate the scene either by physically moving within their play area or using the controller's joystick for locomotion if physical space is limited. Our multiplayer mode can host up to 16 players simultaneously.

\begin{figure}[t!]
    \centering
    \includegraphics[width=\columnwidth]{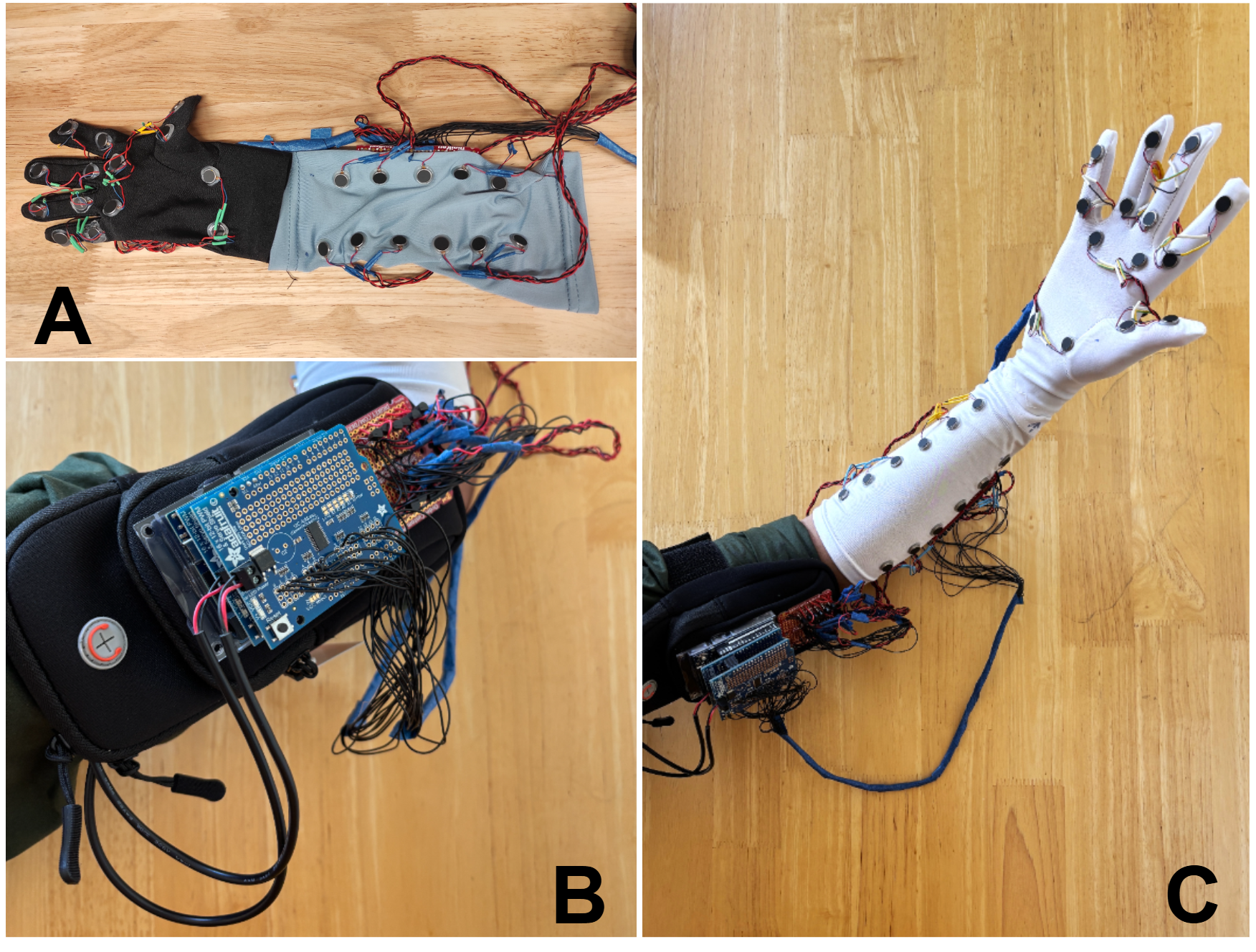}
    \captionsetup{belowskip=-10pt}
    \caption{(A) Glove and sleeve with 26 ERMs, (B) Control unit consisting of 2 stacked PWM shields, an Arduino Uno R4 WiFI, and an integrated power bank, (C) Spatial arrangement of actuators across the hand and forearm when the glove and sleeve are worn}
    \label{fig:hardware}
\end{figure}
\begin{figure*}[t!]
    \centering
    \begin{subfigure}[t]{0.54\textwidth}
        \centering
        \includegraphics[width=\columnwidth]{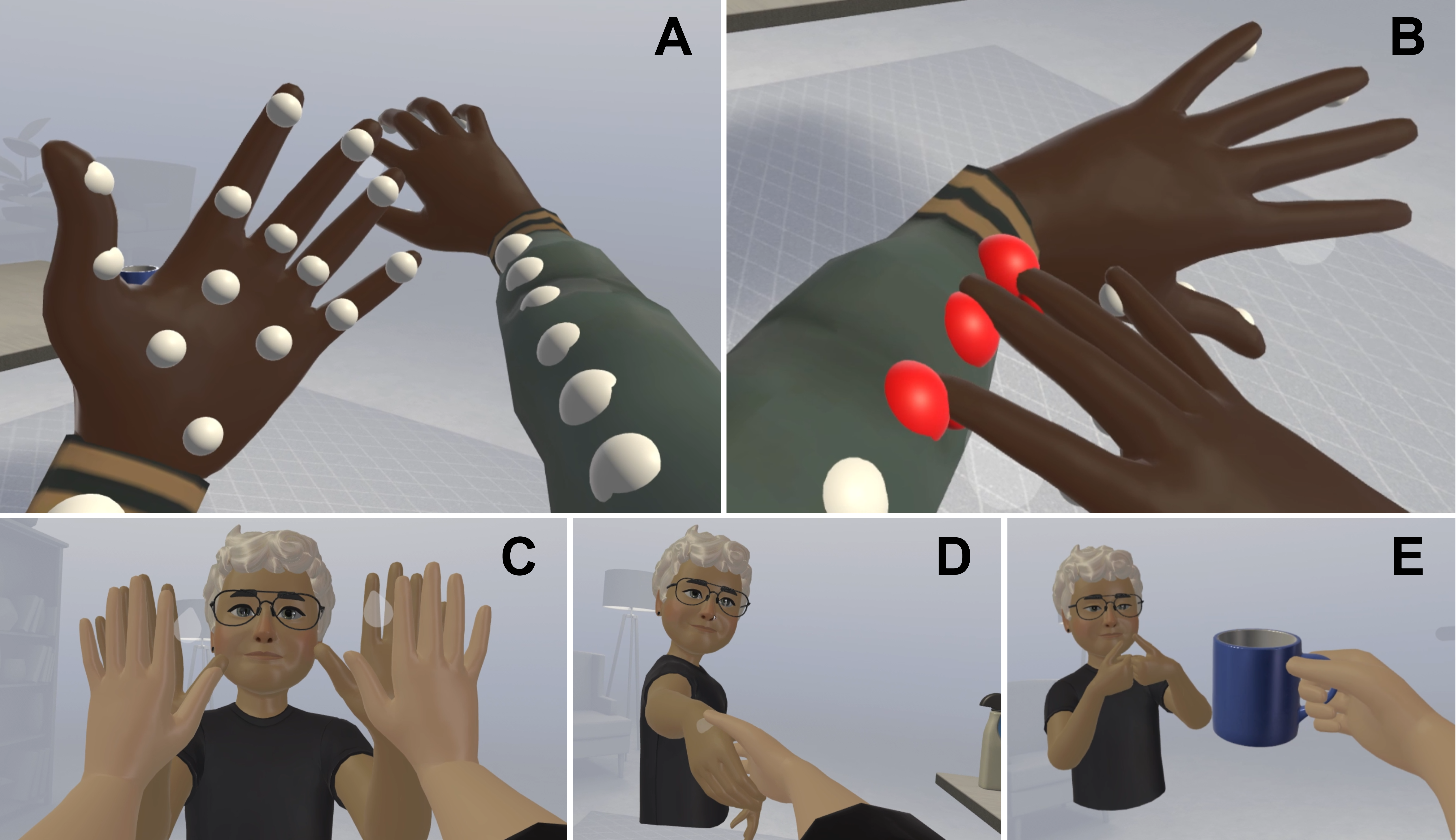}
        \caption{Stills from the VR showing: (A) Spheres attached to the hands and lower arms to detect collision, (B) Visual representation of touch interactions between the hand and forearm. The spheres are not shown to users when they are (C), (D) interacting with each other in VR, and (E) with other virtual objects in the VR environment}
        \label{fig:VRSshots}
    \end{subfigure}
    \hfill
    \begin{subfigure}[t]{0.425\textwidth}
        \centering
        \includegraphics[width=\columnwidth]{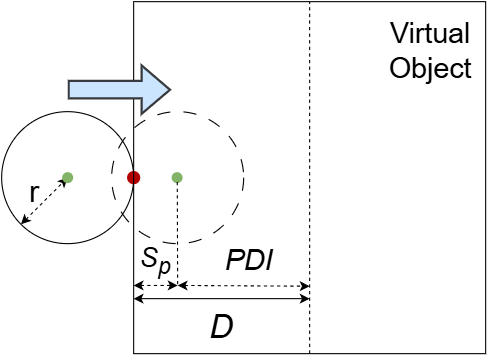}
        \caption{Sphere Colliding with a Virtual Object, where $r$: radius of sphere, $S_p$: distance from point of initial point of contact on surface of object, $D$: maximum penetrable distance, \textit{PDI}: Penetration Depth Information; Green dot: center of sphere, Red dot: first point of collision of sphere with virtual object}
        \label{fig:Collision1}
    \end{subfigure}
    \caption{Visualization of the VR environment interactions and collision dynamics}
    \label{PDI_diagram}
\end{figure*}
\subsection{Collision Information}\label{CollisionInformation}
In our VR setup, avatar collisions are detected using a system of colliders attached to spherical structures that represent joints on the avatar. These spheres, distributed based on the number of actuators in our haptic device, allow for a tailored mapping of tactile feedback. If another device can support a higher tactile resolution/contact points, then there is always an option to add more depending on the number of actuation points of the device.
The developed glove incorporates 14 ERMs across the hand, inspired by the design principles outlined in~\cite{ariza2016inducing}, and 12 ERMs across the lower arm (6 dorsal, 6 volar) inspired by the design principles in~\cite{israrForearm} to ensure continuous, smooth, pleasant motion. These actuators are placed to align with neurological aspects, particularly focusing on the fast-adapting Pacinian corpuscles in the fingers and palm, which are highly responsive to vibrations. Following the approach in \cite{ariza2016inducing}, actuators are distributed to balance device mobility, power consumption, and user comfort, ensuring a functional and unobtrusive wearable device. 

Each finger is equipped with two actuators, one at the fingertip or just above the distal interphalangeal joint, and another just below the proximal interphalangeal joint. Four actuators are placed on the palm, with two below the metacarpophalangeal joints of the index and pinky fingers, one near the carpometacarpal joint, and one just above the wrist joint. Placing a vibration motor in the center of the palm was avoided, as it may not maintain contact with the skin in certain hand postures~\cite{giannopoulos2012touching}.
The forearm contains six actuators each on the dorsal and ventral sides, spanning a 14 cm distance, as configured in~\cite{israrForearm}. There are a total of 26 ERM actuators distributed across each hand (14) and forearm (12), as shown in Figure~\ref{fig:hardware}, creating 52 ERMs in total for each user (left and right hands combined).

We control the ERM actuators using Pulse Width Modulated (PWM) signals, which correspond to the amplitude or intensity of the vibration. These signals were optimized in an initial pilot study involving 8 participants. This optimization aimed to ensure that the vibrations effectively conveyed the desired information while remaining pleasant at the same time. Two key parameters were accounted for: (1) user comfort, as prolonged exposure to continuous vibrations was anticipated, and (2) perceptual variability, ensuring the range was sufficiently wide to allow users to reliably perceive changes in vibration intensity based on the desired information conveyed. 

Our algorithm for detecting and encoding collision information for haptic feedback operates as a two-stage process. In addition to simply identifying the occurrence of a collision (binary yes/no), the algorithm encodes quantitative information about the degree of contact, specifically the amount or distance penetrated into a virtual object. The data sent to each microcontroller consists of a string containing the actuator ID, indicating which actuator should be actuated, alongside collision information that prompts actuation of the specified actuator via the ID. This collision information consists of the Penetration Depth Information (PDI), the contact normal, and a flag that indicates if a collision occurred for the first time. For the remainder of this paper, PDI that is sent over to the microcontroller will be defined as the difference between the maximum possible penetration distance and the distance from the point of initial contact on the surface of the object, as shown in Figure~\ref{fig:Collision1}. 

Each sphere in arrays on the hand and arm is equipped with colliders and tagged to distinguish them from other colliders that do not belong to the spheres specifically attached to the avatar. Upon initial contact with any other collider, we send the collision information to the microcontroller with the PDI set to 0, and set the flag variable to 1 or ``true". This indicates that it is a first-time contact. For our wearable device, the ERM corresponding to the colliding sphere is programmed to produce a short, sharp vibration pulse at the maximum possible amplitude, providing a tactile ``click" sensation that signals the beginning or making of contact. This feature allows users to feel a discrete, high-amplitude pulse upon tapping. If the sphere continues to remain in contact, The flag variable is set to 0 or ``false" immediately after, indicating that it is no longer first contact.



\emph{\textbf{Interacting with Virtual Objects:}} When the user touches a virtual object (which is not a sphere attached to the avatar), if the sphere continues to remain in contact with the collider without pressing further, the PDI is sent as the difference between the maximum possible penetration distance and the distance from the initial contact point on the surface of the object that the sphere collides with. The corresponding ERM begins a continuous vibration but at a reduced amplitude (12\% duty cycle for the hand,  24\% duty cycle for the arm). As the user presses into the collider, the vibration amplitude increases linearly as the PDI decreases. The vibration amplitude is highest (25\% duty cycle for the hand and 50\% for the arm) when the maximum possible penetration distance is reached (i.e. when PDI is 0). Each virtual object in the VR scene has a predefined stiffness, which determines the maximum penetration distance: higher stiffness correlates with a shorter maximum penetrable distance. Once this maximum possible penetration depth is reached, the ERM maintains the highest vibration amplitude, reflecting the resistance of stiffer objects.

If the virtual object is ``grabbable" in the virtual environment and the algorithm detects the user performing a ``grabbing" action, such as picking up the object or holding it in their hand, rather than merely pressing against it while it remains static, the haptic feedback transitions accordingly. Following the initial ``click" sensation, the vibration amplitude is gradually reduced until it reaches a minimal level (12\% duty cycle for the hand). This gradual attenuation of vibrations provides the user with the sensation of holding or grasping the object while mitigating discomfort or unpleasantness caused by prolonged continuous high-intensity vibrations.

\emph{\textbf{Interacting with other Avatars: }} Each avatar has spheres attached to their fingers and forearms, corresponding to the ERMs, which serve as the points of interaction. The radius of the spheres attached to the fingers are smaller in comparison to those across the forearm. On the forearm, larger spheres are used to cover a broader surface area, to ensure smooth, continuous, and consistent vibrations across the skin. In contrast, the smaller spheres on the fingers provide greater precision and accuracy, accommodating the higher density of actuation points required for interactions in a more localized area. When two virtual avatars interact, the collision information between their respective spheres triggers vibrations. The vibration intensity is designed to reflect the extent of their overlap. The constant maximum penetrable distance is set to a constant value, which is the diameter of the smaller sphere involved in the collision.

At the point of initial contact, vibrations are triggered with a minimum amplitude (12\% duty cycle for the hand and 24\% duty cycle for the lower arm). As the overlap increases, the vibration amplitude scales linearly, reaching its maximum when the smaller sphere penetrates another smaller or larger sphere by a depth equal to its own diameter. This design ensures that when two small spheres collide, the maximum amplitude (24\% duty cycle) is achieved when the spheres are completely overlapping. When a larger sphere (attached to the lower arm) interacts with a smaller sphere, or when two larger spheres interact, maximum vibration amplitude (50\% duty cycle) is reached when the smaller sphere is fully enclosed within the larger one or when the larger spheres overlap each other by a distance equal to the smaller sphere's diameter.

When no collision happens, no information is sent over to the microcontroller, and the flag variable is reset to 1 or ``true", indicating that whenever the next collision happens, it will again collide for the first time.

\subsection{Hardware Setup}

The system is designed to be wireless and plug-and-play, making it adaptable to any tactile device that supports WiFi connectivity and/or serial communication. For controlling the wearables that we developed, we employed the Arduino Uno R4 WiFi microcontroller, with two separate microcontrollers connected to the same WiFi network as the VR headset -- one designated for the left hand and the other for the right hand. For each hand, a user is equipped with a haptic glove, a forearm sleeve, an Arduino Uno R4 WiFi module, two 16-Channel 12-bit PWM shields (Adafruit, PCA9685) that are stacked on top of each other on top of the Arduino (Figure~\ref{fig:hardware}), a 5V 10000mAh power bank that is capable of powering the glove, sleeve, Arduino, and the PWM shields, and a running armband that securely houses the power bank along with all associated electronic components. This makes our hardware easily wearable and portable, without needing to be connected to an external computer or power source. Furthermore, the gloves and sleeves we use are made of quality spandex, a soft and smooth material that is comfortable to wear, designed to be breathable, and fits a large variety of hand sizes.

Each microcontroller together with the PWM shields is configured to drive simple 11000 RPM 5V vibration motors or ERMs (Adafruit, 1528-1177-ND) integrated into the hand gloves and armbands. ERMs were chosen for their low latency, small form factor, and simplicity of operation in real-time applications. These qualities allow for VR compatibility, ease of integration into wearable designs, and perceptible haptic feedback within a usable range of frequencies and amplitudes. Each ERM is connected to the PWM shield via a transistor, which controls the amount of current flow to generate a wide range of vibrotactile sensations. 

\subsection{Wireless Communication}
To identify and communicate with each device over WiFi, each microcontroller sends an initial User Datagram Packet (UDP) containing its identifier, i.e. ``left” or ``right” hand, along with its local IP address. The Unity application initiates the connection process by broadcasting a message and receiving callbacks from both microcontrollers. Once local IP addresses are obtained, Unity establishes a stable, dedicated connection with each microcontroller for each hand using the Transmission Control Protocol (TCP). TCP was chosen over UDP to ensure reliable packet delivery.

The information sent is maximum 21 bytes, which is well below the Arduino's Maximum Transmission Unit (MTU) limits, ensuring minimal impact on transmission speed and latency. Therefore, if needed, we can append and transmit more information than our algorithm currently does. The PDI is updated in every Unity frame which is set to run at 120 FPS, and the frequency of the entire system (including the refresh rate of the microcontroller) ranges from 100-120 Hz. All calculations of amplitude and frequency are processed on the Arduino. 
Although amplitude and frequency values could be sent directly, the choice of transmitting collision data provides greater flexibility and universality. Most importantly, this approach enables compatibility with a range of devices, as the information is encoded and sent in such a way that each device can decode the collision data according to its unique actuation mechanism. For example, devices capable of rendering force or pressure may interpret collision data differently than those using voice coil or ultrasonic actuation, ensuring a versatile, device-agnostic solution. In our case we do not make use of the contact normals that are sent, as our device only provides vibrotactile feedback, but it can be useful for devices that provide pressure or force feedback. 

\section{Experimental Methods}

We developed four haptic glove and sleeve prototypes to facilitate bilateral interactions between two users and provide haptic feedback to both users on each hand and forearm (left and right). 

We conducted a two-part user study with experiments involving two participants, each located in separate physical spaces. The order of the two experiments in this study design was counterbalanced across participant pairs to mitigate order effects.
\textit{Part 1: Prescribed Touch} - Participants engaged in a structured task where the toucher performed specific touch gestures on the touchee as directed by the experimenter. 
\textit{Part 2: Free-Form Interaction} - In this exploratory task, participants interacted freely with each other and with other virtual objects within the VR environment.

\subsection{Participants}
In this study, pairs of individuals participated in the study together. 
Since people had to interact through touch and perform gestures such as stroking, poking, patting and squeezing, we recruited pairs of participants who had a current relationship (i.e., friends, family, or romantic partners)~\cite{thompson2011effect}. 
In total, 12 participants (6 female, 5 male, 1 non-binary, $M_{age} = 25 \pm 3.22$), were recruited for the study, of which 2 had prior experience with haptic devices and VR. None of the participants had sensory or motor impairments. The study was approved by the University XX Institutional Review Board under protocol XXXX, and all participants gave informed consent. Participants wore noise-cancelling earphones connected to the VR headset, which mitigated noise from the ERMs; the earphones also delivered audio from the VR simulation for Experiment 2 only. To eliminate potential auditory bias, audio feedback was intentionally removed for Experiment 1.

\subsection{Procedure}
Experiments 1 and 2 were evenly balanced across all pairs of participants to eliminate any potential order effects. Half of the pairs of participants performed Experiment 1 first, followed by Experiment 2, while the remaining half performed Experiment 2 first, followed by Experiment 1. In both the experiments, we utilized two avatars from the Meta Avatar SDK pack. The exploration of variables such as appearance, gender, skin tone, facial expressions, or other avatar characteristics is beyond the scope of this study. To minimize bias or personalization, participants were informed that the avatars might not resemble their actual selves.

\subsubsection{\textbf{Experiment 1 -- Prescribed Touch}} 
This experiment assessed the ability of our VR application, in combination with the developed haptic glove and sleeve, to facilitate the remote communication of social touch gestures. The experiment involved a structured task where the toucher, guided by the experimenter, performed four predefined gestures - stroke, pat, poke, and squeeze~\cite{hertenstein2009communication} - on the touchee. A within-subjects experiment was conducted, wherein participants alternated roles as the toucher and the touchee. Participants were asked to complete the gestures at three distinct speeds: slow (S), medium (M), and fast (F). Participants performed these gestures under two feedback conditions: (1) visual feedback only (V) and (2) combined visual and haptic feedback (VH). The 24 trials (4 gestures $\times$ 3 speeds $\times$ 2 feedback) were counterbalanced using a Graeco-Latin square design to preclude any potential order effects. Consequently, a participant pair performed 48 trials (24 $\times$ 2 user modes) in total. During each trial, participants were instructed to perform the gesture three times, and were instructed to not verbally communicate with each other in VR. 

Before starting the experiment, each toucher underwent a training phase of approximately 5 minutes, during which they were shown and asked to practice performing poke, pat, squeeze and stroke gestures on a static avatar's arm, receiving visual feedback only (no haptic feedback). The medium (M) speed was chosen as the normal speed at which users would typically perform the gesture. For the other two speeds, slow (S) and fast (F), users were asked to perform the gesture at a speed significantly slower and faster than the medium (M) speed. Only during the training phase were the real-time hand velocities and mean velocity for each gesture performed as visual feedback to all participants. This was intended to help them better assess the relative speed of their movements and maintain consistency throughout the experiment.

A 10-minute break was provided when participants exchanged roles as the toucher and touchee. Thus, each participant pair completed the entire experiment in approximately 1 hour and 30 minutes.

\emph{\textbf{Experimental Measures:}}
To investigate the relationship between participants' real-world comfort with interpersonal touch and their experience of virtual social body contact, we administered the Comfort with Interpersonal Touch (CIT) questionnaire~\cite{webb2015individual} prior to the experiment. 
To monitor potential cybersickness, participants also completed the Simulator Sickness Questionnaire (SSQ)~\cite{kennedy1993simulator} both before and after the experiment.

During the experiment, we collected both objective and subjective data. From the toucher, we recorded the real-time trajectory of the hand (3D position) and the velocity within the VR environment. We also recorded the different speeds at which the toucher performed the gestures. For stroking, we calculated the mean velocity at which the toucher performed the gesture in $cm/s$. For squeeze, poke and pat, we measured the time taken to perform each gesture 3 times for each speed and each feedback condition, and then averaged them to calculate the number of squeezes/second, pokes/second, or pats/second. 

After initiating/receiving each gesture, both toucher and touchee provided feedback through the following subjective measures in VR: (a) Emotional responses -- Valence and Arousal -- on a 2D Valence-Arousal Emojigrid ranging from -1 to +1~\cite{toet2019emojigrid}, (b) \textbf{S1 (Plausibility\footnote{Degree to which the haptic feedback aligns with user's expectations of how something should feel, in relation to visual cues or the user's prior experience}):} \textit{The touch sensation I felt in my body corresponded to the virtual touch I saw}, (c) \textbf{S2 (Pleasantness):} \textit{How `pleasant' did the interaction feel?}, and (d) \textbf{S3 (Naturalness):} \textit{How `natural' did the interaction feel?} Ratings for (b), (c) and (d) were provided on a 7-point Likert scale, where 1 = \textit{not at all} to 7 = \textit{very much so}. 

\begin{figure}
    \centering
    \includegraphics[width=\columnwidth]{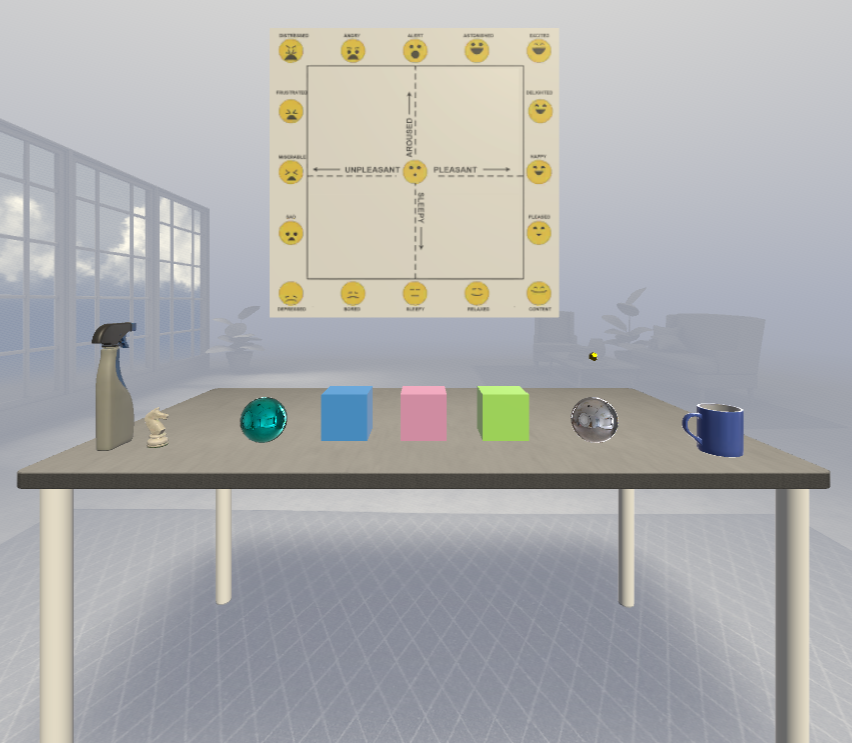}
    \captionsetup{belowskip = -15pt}
    \caption{Interactable virtual objects and 2D Valence-Arousal Emojigrid~\cite{toet2019emojigrid}}
    \label{fig:enter-label}
\end{figure}
To evaluate participants’ sense of presence in the VR environment, we administered the Slater-Usoh-Steed (SUS) questionnaire~\cite{slater1998,usoh1999}, rephrased for VR, at the end of the experiment. 
Additionally, first-time touchers (participants who had not previously been touchees) answered an open-ended question: \textit{``What visual indications did you compare the performed touch gestures to?”}

\subsubsection{\textbf{Experiment 2 -- Free Interaction}}
Participants were instructed on the proper usage of the gloves and sleeves, as well as the procedure for launching the VR application on the Meta Quest headset. Following this, they were asked to independently don the hardware and initiate the VR application without further assistance. After launching the VR application, participants were asked to explore and interact freely within the VR environment, the only condition being that they find a way to interact with each other through touch using their hands or lower arms. If direct interaction felt uncomfortable or unnatural, participants were encouraged to create a secret handshake that involved their hands and/or lower arms. Beyond this requirement, participants were free to interact with other virtual objects or collaboratively perform tasks of their choice. The experiment was conducted under two feedback conditions, counterbalanced to preclude any order effects: (1) visual feedback only (V) and (2) combined visual and haptic feedback (VH). Each free exploration session lasted 10 minutes, with a 5-minute break between feedback conditions. The total duration of the experiment was 25 minutes. Participants were allowed to verbally communicate with each other in VR during the task.

\emph{\textbf{Experimental Measures:}}
We used the Embodiment Questionnaire~\cite{peck2021avatar} to measure how strongly participants felt connected to their virtual avatars in the VR system, the SUS questionnaire, rephrased for VR, to assess presence, and the SSQ both pre- and post-experiment to monitor cybersickness. To evaluate the perceived usability of our VR system, participants were asked to complete the System Usability Scale questionnaire~\cite{brooke1996sus}. All of these questionnaires were administered for both the V and VH conditions. After the VH condition, participants documented their haptic experience (HX) using the 5-point Likert scale questionnaire from~\cite{anwar2023factors}.

\section{Results}
\begin{figure*}[t!]
    \centering
    \includegraphics[width=\textwidth]{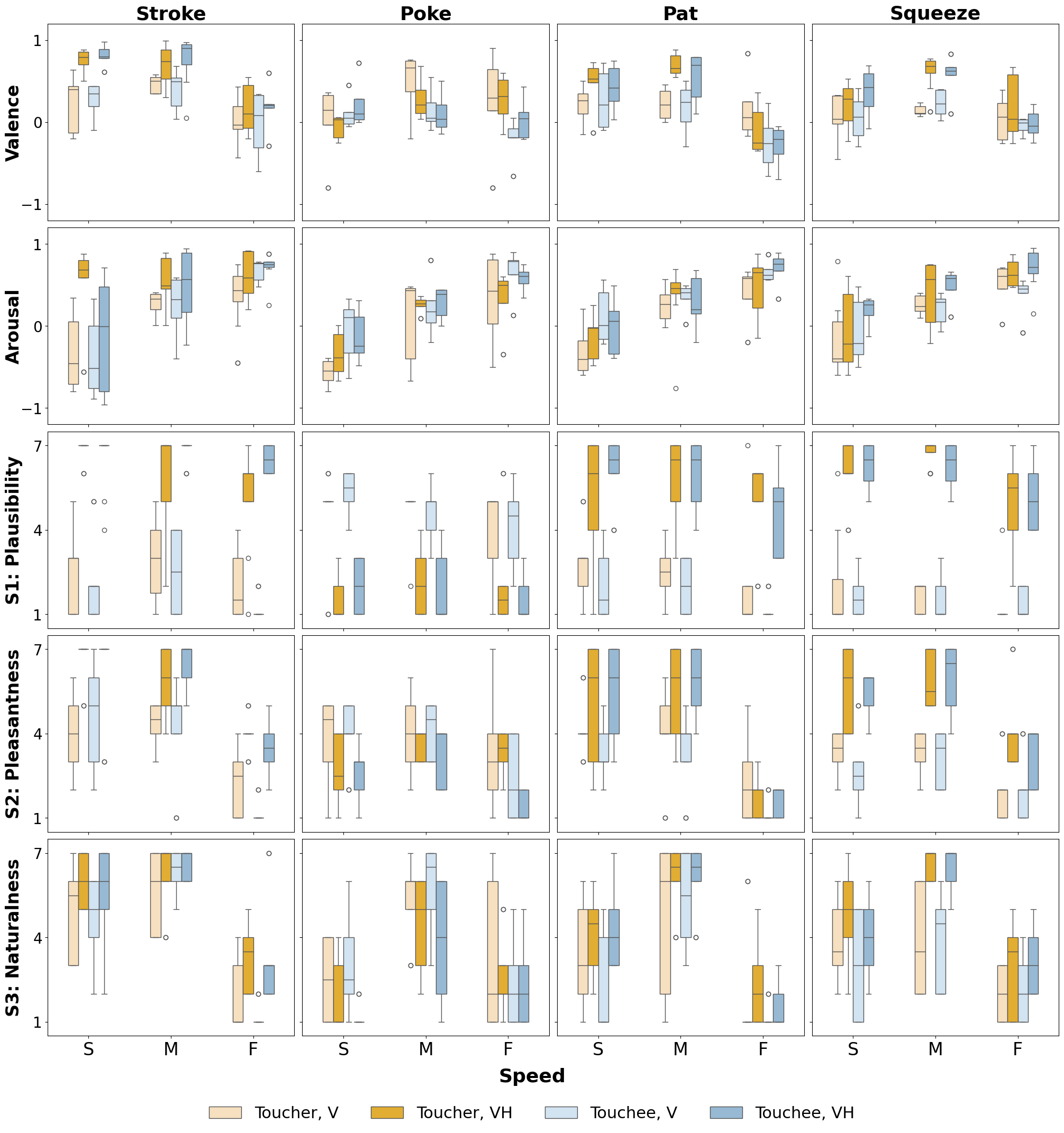}
    \caption{Prescribed Touch Experiment: Ratings of both toucher and touchee across different speeds, user modes, and feedback conditions for each gesture performed}
    \label{}
\end{figure*}

No order effects were observed across Experiments 1 and 2, the results were consistent regardless of whether participants completed Experiment 1 or Experiment 2 first.

\subsection{Experiment 1 -- Prescribed Touch}
\subsubsection{Subjective Measures} For each gesture (Stroke, Poke, Pat, Squeeze) that was performed, we conducted separate 3-way ANOVAs to analyze the effect of Speed (S, M, F), Feedback Modality (V, VH), and User Mode (\textit{Toucher}, \textit{Touchee}) on the \textit{Valence}, \textit{Arousal}, \textit{S1} (Plausibility), 
\textit{S2} (Pleasantness) and \textit{S3} (Naturalness).

\begin{table*}[]
\caption{Tukey's HSD Post Hoc Analysis Results across the different Speed conditions for each Gesture performed}
\label{TablePostHoc}
\renewcommand{\arraystretch}{1.7}
\resizebox{\textwidth}{!}{%
\fontsize{20}{20}\selectfont
\begin{tabular}{|c|ccc|ccc|ccc|ccc|}
\hline
\multirow{2}{*}{\textbf{\begin{tabular}[c]{@{}c@{}}Dependent\\ Variables\end{tabular}}} & \multicolumn{3}{c|}{\textit{\textbf{Stroke}}}                                                                                                                                                                                                                         & \multicolumn{3}{c|}{\textit{\textbf{Poke}}}                                                                                                                                                                                                                          & \multicolumn{3}{c|}{\textit{\textbf{Pat}}}                                                                                                                                                                                                                           & \multicolumn{3}{c|}{\textit{\textbf{Squeeze}}}                                                                                                                                                                                                                      \\[10pt] \cline{2-13} 
                                                                                        & \multicolumn{1}{c|}{\textbf{S vs M}}                                                         & \multicolumn{1}{c|}{\textbf{M vs F}}                                                         & \textbf{F vs S}                                                         & \multicolumn{1}{c|}{\textbf{S vs M}}                                                        & \multicolumn{1}{c|}{\textbf{M vs F}}                                                         & \textbf{F vs S}                                                         & \multicolumn{1}{c|}{\textbf{S vs M}}                                                        & \multicolumn{1}{c|}{\textbf{M vs F}}                                                         & \textbf{F vs S}                                                         & \multicolumn{1}{c|}{\textbf{S vs M}}                                                        & \multicolumn{1}{c|}{\textbf{M vs F}}                                                         & \textbf{F vs S}                                                        \\[10pt]\hline
Valence                                                                                 & \multicolumn{1}{c|}{$p > 0.05$}                                                      & \multicolumn{1}{c|}{\begin{tabular}[c]{@{}c@{}}$\Delta M = -0.5$ \\ $(p < 0.001)$\end{tabular}}  & \begin{tabular}[c]{@{}c@{}}$\Delta M = 0.43$ \\ $(p < 0.001)$\end{tabular}  & \multicolumn{1}{c|}{\begin{tabular}[c]{@{}c@{}}$\Delta M = 0.18$ \\ $(p = 0.02)$\end{tabular}}          & \multicolumn{1}{c|}{\begin{tabular}[c]{@{}c@{}}$\Delta M = -0.16$ \\ $(p = 0.045)$\end{tabular}}         & $p > 0.05$                                                      & \multicolumn{1}{c|}{$p > 0.05$}                                                     & \multicolumn{1}{c|}{\begin{tabular}[c]{@{}c@{}}$\Delta M = -0.53$ \\ $(p < 0.001)$\end{tabular}} & \begin{tabular}[c]{@{}c@{}}$\Delta M = 0.47$ \\ $(p 0< .001)$\end{tabular}  & \multicolumn{1}{c|}{\begin{tabular}[c]{@{}c@{}}$\Delta M = 0.21$ \\ $(p < 0.001)$\end{tabular}} & \multicolumn{1}{c|}{\begin{tabular}[c]{@{}c@{}}$\Delta M = -0.35$ \\ $(p < 0.001)$\end{tabular}} & \begin{tabular}[c]{@{}c@{}}$\Delta M = 0.14$ \\ $(p = 0.036)$\end{tabular}         \\[10pt] \hline
Arousal                                                                                 & \multicolumn{1}{c|}{\begin{tabular}[c]{@{}c@{}}$\Delta M = -0.46$ \\ $(p < 0.001)$\end{tabular}} & \multicolumn{1}{c|}{$p > 0.05$}                                                      & \begin{tabular}[c]{@{}c@{}}$\Delta M = -0.66$ \\ $(p < 0.001)$\end{tabular} & \multicolumn{1}{c|}{\begin{tabular}[c]{@{}c@{}}$\Delta M = 0.49$ \\ $(p < 0.001)$\end{tabular}} & \multicolumn{1}{c|}{\begin{tabular}[c]{@{}c@{}}$\Delta M = 0.27$ \\ $(p < 0.001)$\end{tabular}}  & \begin{tabular}[c]{@{}c@{}}$\Delta M = -0.76$ \\ $(p < 0.001)$\end{tabular} & \multicolumn{1}{c|}{\begin{tabular}[c]{@{}c@{}}$\Delta M = 0.4$ \\ $(p < 0.001)$\end{tabular}}  & \multicolumn{1}{c|}{\begin{tabular}[c]{@{}c@{}}$\Delta M = 0.26$ \\ $(p < 0.001)$\end{tabular}}  & \begin{tabular}[c]{@{}c@{}}$\Delta M = -0.66$ \\ $(p < 0.001)$\end{tabular} & \multicolumn{1}{c|}{\begin{tabular}[c]{@{}c@{}}$\Delta M = 0.39$ \\ $(p < 0.001)$\end{tabular}} & \multicolumn{1}{c|}{\begin{tabular}[c]{@{}c@{}}$\Delta M = -0.22$ \\ $(p = 0.002)$\end{tabular}}         & \begin{tabular}[c]{@{}c@{}}$\Delta M = -0.6$ \\ $(p < 0.001)$\end{tabular} \\[10pt] \hline
Plausibility                                                                                      & \multicolumn{1}{c|}{$p > 0.05$}                                                      & \multicolumn{1}{c|}{$p > 0.05$}                                                      & $p > 0.05$                                                      & \multicolumn{1}{c|}{$p > 0.05$}                                                     & \multicolumn{1}{c|}{$p > 0.05$}                                                      & $p > 0.05$                                                      & \multicolumn{1}{c|}{$p > 0.05$}                                                     & \multicolumn{1}{c|}{$p > 0.05$}                                                      & $p > 0.05$                                                      & \multicolumn{1}{c|}{$p > 0.05$}                                                     & \multicolumn{1}{c|}{$p > 0.05$}                                                      & $p > 0.05$                                                     \\[10pt] \hline
Pleasantness                                                                                      & \multicolumn{1}{c|}{$p > 0.05$}                                                      & \multicolumn{1}{c|}{\begin{tabular}[c]{@{}c@{}}$\Delta M = -2.5$ \\ $(p < 0.001)$\end{tabular}}  & \begin{tabular}[c]{@{}c@{}}$\Delta M = 2.67$ \\ $(p < 0.001)$\end{tabular}  & \multicolumn{1}{c|}{$p > 0.05$}                                                     & \multicolumn{1}{c|}{\begin{tabular}[c]{@{}c@{}}$\Delta M = -1.21$ \\ $(p < 0.001)$\end{tabular}} & \begin{tabular}[c]{@{}c@{}}$\Delta M = 0.71$ \\ $(p = 0.023)$\end{tabular}          & \multicolumn{1}{c|}{$p > 0.05$}                                                     & \multicolumn{1}{c|}{\begin{tabular}[c]{@{}c@{}}$\Delta M = -2.96$ \\ $(p < 0.001)$\end{tabular}} & \begin{tabular}[c]{@{}c@{}}$\Delta M = 2.79$ \\ $(p < 0.001)$\end{tabular}  & \multicolumn{1}{c|}{$p > 0.05$}                                                     & \multicolumn{1}{c|}{\begin{tabular}[c]{@{}c@{}}$\Delta M = -1.71$ \\ $(p < 0.001)$\end{tabular}} & \begin{tabular}[c]{@{}c@{}}$\Delta M = 1.42$ \\ $(p < 0.001)$\end{tabular} \\[10pt] \hline
Naturalness                                                                                      & \multicolumn{1}{c|}{\begin{tabular}[c]{@{}c@{}}$\Delta M = 0.96$ \\ $(p = 0.003)$\end{tabular}}          & \multicolumn{1}{c|}{\begin{tabular}[c]{@{}c@{}}$\Delta M = -3.83$ \\ $(p < 0.001)$\end{tabular}} & \begin{tabular}[c]{@{}c@{}}$\Delta M = 2.88$ \\ $(p < 0.001)$\end{tabular}  & \multicolumn{1}{c|}{\begin{tabular}[c]{@{}c@{}}$\Delta M = 2.79$ \\ $(p < 0.001)$\end{tabular}} & \multicolumn{1}{c|}{\begin{tabular}[c]{@{}c@{}}$\Delta M = -2.33$ \\ $(p < 0.001)$\end{tabular}} & $p > 0.05$                                                      & \multicolumn{1}{c|}{\begin{tabular}[c]{@{}c@{}}$\Delta M = 2.12$ \\ $(p < 0.001)$\end{tabular}} & \multicolumn{1}{c|}{\begin{tabular}[c]{@{}c@{}}$\Delta M = -3.92$ \\ $(p < 0.001)$\end{tabular}} & \begin{tabular}[c]{@{}c@{}}$\Delta M = 1.79$ \\ $(p < 0.001)$\end{tabular}  & \multicolumn{1}{c|}{\begin{tabular}[c]{@{}c@{}}$\Delta M = 1.33$ \\ $(p < 0.001)$\end{tabular}} & \multicolumn{1}{c|}{\begin{tabular}[c]{@{}c@{}}$\Delta M = -2.67$ \\ $(p < 0.001)$\end{tabular}} & \begin{tabular}[c]{@{}c@{}}$\Delta M = 1.33$ \\ $(p < 0.001)$\end{tabular} \\[10pt] \hline
\end{tabular}
}
\end{table*}

\textbf{Stroke: }Speed had a main effect on \textit{Valence} ($F(2, 132) = 58.78, p < 0.001, \eta_p^2 = 0.25$), \textit{Arousal} ($F(2, 132) = 36.44, p < 0.001, \eta_p^2 = 0.19$), \textit{Plausibility} ($F(2, 132) = 6.97, p = 0.001, \eta_p^2 = 0.02)$, \textit{Pleasantness} ($F(2, 132) = 77.97, p < 0.001, \eta_p^2 = 0.29$), and \textit{Naturalness} ($F(2, 132) = 116.20, p < 0.001, \eta_p^2 = 0.44$). 
Feedback modality had a main effect on \textit{Valence} ($F(1, 132) = 65.17, p < 0.001, \eta_p^2 = 0.14$), \textit{Arousal} ($F(1, 132) = 24.24, p < 0.001, \eta_p^2 = 0.06$), \textit{Plausibility} ($F(1, 132) = 404.59, p < 0.001, \eta_p^2 = 0.58$), \textit{Pleasantness} ($F(1, 132) = 104.82, p < 0.001, \eta_p^2 = 0.20$), and \textit{Naturalness} ($F(1, 132) = 25.70, p < 0.001, \eta_p^2 = 0.05$). 
User Mode had no significant main effects on any dependent variables ($p > 0.05$), except for a significant interaction effect with Feedback on \textit{Plausibility} ($F(1, 132) = 9.06, p = 0.003, \eta_p^2 = 0.01$) and \textit{Arousal} ($F(1,132) = 4.45, p = 0.04, \eta_p^2 = 0.01$), and with Speed on \textit{Arousal} ($F(2, 132) = 6.42, p = 0.002, \eta_p^2 = 0.03$) and \textit{Pleasantness} ($F(2, 132) = 3.48, p = 0.033, \eta_p^2 = 0.01$).
Significant interaction effects were observed for Speed and Feedback on \textit{Valence} ($F(2, 132) = 6.54, p = 0.002, \eta_p^2 = 0.03$), \textit{Arousal} ($F(2, 132) = 3.92, p = 0.022, \eta_p^2 = 0.02$), and \textit{Naturalness} ($F(2,132) = 3.40, p = 0.04, \eta_p^2 = 0.01$). Tukey's HSD posthoc results are shown in Table~\ref{TablePostHoc}.

\textbf{Poke: }Speed had a main effect on \textit{Valence} ($F(2, 132) = 5.39, p = 0.006, \eta_p^2 = 0.03$), \textit{Arousal} ($F(2, 132) = 75.88, p < 0.001, \eta_p^2 = 0.34$), \textit{Pleasantness} ($F(2, 132) = 12.93, p < 0.001, \eta_p^2 = 0.08$), and \textit{Naturalness} ($F(2, 132) = 43.08, p < 0.001, \eta_p^2 = 0.23$).
Feedback modality had a main effect on \textit{Plausibility} ($F(1, 132) = 173.54, p < 0.001, \eta_p^2 = 0.38$), \textit{Pleasantness} ($F(1, 132) = 15.91, p < 0.001, \eta_p^2 = 0.05$), and \textit{Naturalness} ($F(1, 132) = 15.23, p < 0.001, \eta_p^2 = 0.04$). 
User Mode had a main effect on \textit{Valence} ($F(1, 132) = 9.78, p = 0.002, \eta_p^2 = 0.03$), \textit{Arousal} ($F(1, 132) = 21.68, p < 0.001, \eta_p^2 = 0.05$), and \textit{Pleasantness} ($F(1, 132) = 6.58, p = 0.011, \eta_p^2 = 0.02$).
Significant interaction effects were observed between Speed and User Mode on \textit{Valence} ($F(2, 132) = 10.58, p < 0.001, \eta_p^2 = 0.07$) and \textit{Pleasantness} ($F(2, 132) = 6.64, p = 0.002, \eta_p^2 = 0.04$) only. Tukey's HSD posthoc results are shown in Table~\ref{TablePostHoc}.

\textbf{Pat:}
Speed had a main effect on \textit{Valence} ($F(2, 132) = 56.89, p < 0.001, \eta_p^2 = 0.26$), \textit{Arousal} ($F(2, 132) = 70.66, p < 0.001, \eta_p^2 = 0.33$), \textit{Plausibility} ($F(2, 132) = 5.58, p = 0.005, \eta_p^2 = 0.02$), \textit{Pleasantness} ($F(2, 132) = 80.53, p < 0.001, \eta_p^2 = 0.34$), and \textit{Naturalness} ($F(2, 132) = 76.76, p < 0.001, \eta_p^2 = 0.35$). 
Feedback modality had a main effect on \textit{Valence} ($F(1, 132) = 13.50, p < 0.001, \eta_p^2 = 0.03$), \textit{Plausibility} ($F(1, 132) = 228.86, p < 0.001, \eta_p^2 = 0.45$), \textit{Pleasantness} ($F(1, 132) = 29.76, p < 0.001, \eta_p^2 = 0.06$), and \textit{Naturalness} ($F(1, 132) = 14.97, p < 0.001, \eta_p^2 = 0.03$), respectively. 
User Mode had a main effect on \textit{Valence} ($F(1, 132) = 7.72, p = 0.006, \eta_p^2 = 0.02$) and \textit{Arousal} ($F(1, 132) = 11.93, p < 0.001, \eta_p^2 = 0.03$).
Significant interaction effects were observed for Speed and Feedback on \textit{Valence} ($F(2, 132) = 14.71, p < 0.001, \eta_p^2 = 0.07$) and \textit{Pleasantness} ($F(2, 132) = 7.64, p < 0.001, \eta_p^2 = 0.03$); for Speed and User Mode on \textit{Valence} ($F(2, 132) = 3.60, p = 0.030, \eta_p^2 = 0.02$) and \textit{Arousal} ($F(2, 132) = 3.55, p = 0.031, \eta_p^2 = 0.02$); and for Feedback and User Mode on \textit{Plausibility} ($F(1, 132) = 4.23, p = 0.042, \eta_p^2 = 0.01$) and \textit{Pleasantness} ($F(1, 132) = 5.47, p = 0.021, \eta_p^2 = 0.01$). Tukey's HSD posthoc results are shown in Table~\ref{TablePostHoc}.

\textbf{Squeeze:}
Speed had a main effect on \textit{Valence} ($F(2, 132) = 27.25, p < 0.001, \eta_p^2 = 0.14$), \textit{Arousal} ($F(2, 132) = 56.67, p < 0.001, \eta_p^2 = 0.28$), \textit{Plausibility} ($F(2, 132) = 11.16, p < 0.001, \eta_p^2 = 0.02$), \textit{Pleasantness} ($F(2, 132) = 35.43, p < 0.001, \eta_p^2 = 0.14$), and \textit{Naturalness} ($F(2, 132) = 45.06, p < 0.001, \eta_p^2 = 0.22$).
Feedback modality had a main effect on \textit{Valence} ($F(1, 132) = 37.74, p < 0.001, \eta_p^2 = 0.10$), \textit{Arousal} ($F(1, 132) = 21.18, p < 0.001, \eta_p^2 = 0.05$), \textit{Plausibility} ($F(1, 132) = 722.07, p < 0.001, \eta_p^2 = 0.71$), \textit{Pleasantness} ($F(1, 132) = 173.25, p < 0.001, \eta_p^2 = 0.33$), and \textit{Naturalness} ($F(1, 132) = 45.99, p < 0.001, \eta_p^2 = 0.11$).
User Mode had no significant main effects on any dependent variables ($p > 0.05$). Significant interaction effects were observed for Speed and Feedback on \textit{Valence}, $F(2, 132) = 6.18, p = 0.003, \eta_p^2 = 0.03$, \textit{Plausibility}, $F(2, 132) = 3.72, p = 0.027, \eta_p^2 = 0.01$, and \textit{Naturalness}, $F(2, 132) = 5.87, p = 0.004, \eta_p^2 = 0.03$ only. Tukey's HSD posthoc results are shown in Table~\ref{TablePostHoc}.

\begin{figure}[t!]
    \centering
    \includegraphics[width=\columnwidth]{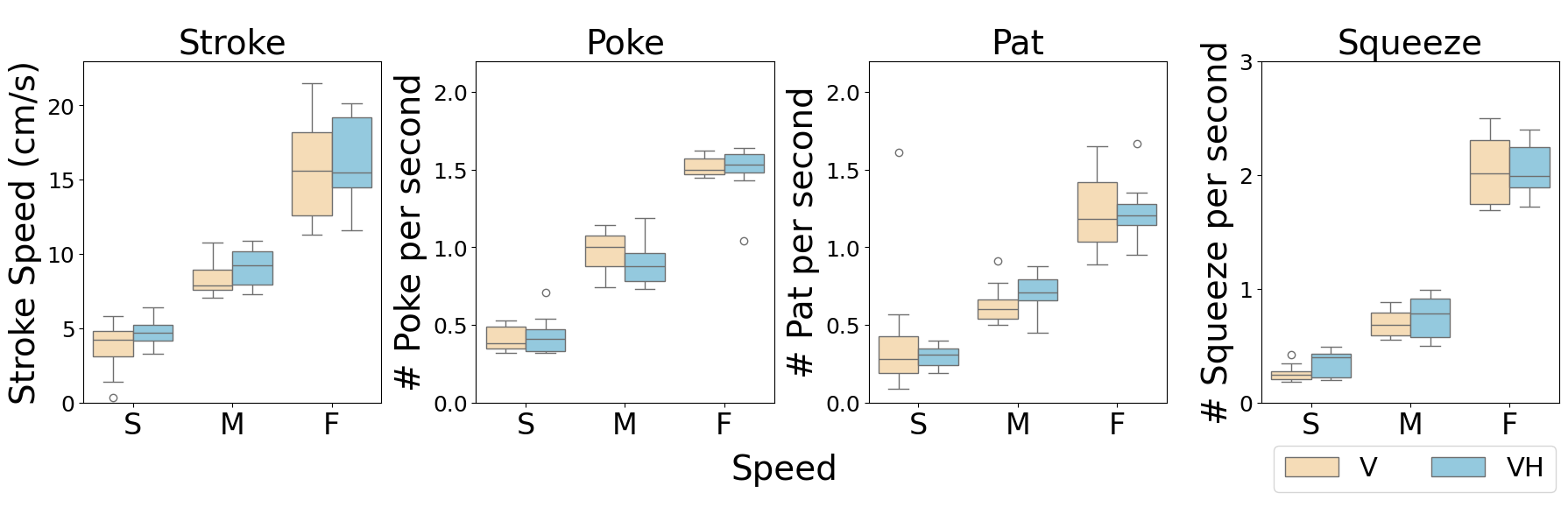}
    \captionsetup{belowskip=-15pt}
    \caption{Recorded gesture speed performed by Toucher, categorized into the three prescribed speed levels}
    \label{recordedSpeed}
\end{figure}
\subsubsection{Objective Measures During Experiment}

Objective measures of the interaction -- stroking speed and the frequency of poke, pat, and squeeze actions per second -- varied significantly between participants (Figure~\ref{recordedSpeed}), corresponding to the 3 speed conditions (S, M, F). To assess whether these objective measures significantly influenced participant ratings of \textit{Valence}, \textit{Arousal}, \textit{Plausibility}, \textit{Pleasantness} and \textit{Naturalness}, we conducted Pearson correlation analyses separately for both the \textit{Toucher} and the \textit{Touchee}.

\textbf{Stroke: }
For the \textit{Toucher}, there was a negative correlation between stroking speed and Valence ($r(70) = -0.44$, $p < 0.001$), Pleasantness ($r(70) = -0.50$, $p < 0.001$), and Naturalness ($r(70) = -0.59$, $p < 0.001$), and a positive correlation with Arousal ($r(70) = 0.43$, $p < 0.001$). No significant correlation was observed for Plausibility ($p > 0.05$).
For the \textit{Touchee}, a similar pattern was found with negative correlations between stroking speed and Valence ($r(70) = -0.46$, $p < 0.001$), Pleasantness ($r(70) = -0.52$, $p < 0.001$), and Naturalness ($r(70) = -0.49$, $p < 0.001$), and a positive correlation with Arousal ($r(70) = 0.60$, $p < 0.001$). No significant correlation was observed for Plausibility ($p > 0.05$).

\textbf{Poke: }
For the \textit{Toucher}, there was a positive correlation between number of pokes per second and Arousal ($r(70) = 0.67$, $p < 0.001$) and Valence ($r(70) = 0.28$, $p < 0.05$). No significant correlation was observed for Plausibility, Pleasantness, or Naturalness ($p > 0.05$).
For the \textit{Touchee}, there was a strong positive correlation between number of pokes per second and Arousal ($r(70) = 0.82$, $p < 0.001$) and a moderate negative correlation with Pleasantness ($r(70) = -0.44$, $p < 0.001$). No significant correlation was observed for Valence, Plausibility, or Naturalness ($p > 0.05$).

\textbf{Pat: }
For the \textit{Toucher}, there was a negative correlation between number of pats per second and Valence ($r(70) = -0.39$, $p < 0.001$), Pleasantness ($r(70) = -0.56$, $p < 0.001$), and Naturalness ($r(70) = -0.33$, $p < 0.01$), and a positive correlation with Arousal ($r(70) = 0.52$, $p < 0.001$). No significant correlation was observed for Plausibility ($p > 0.05$).
For the \textit{Touchee}, there was a negative correlation between number of pats per second and Valence ($r(70) = -0.49$, $p < 0.001$), Pleasantness ($r(70) = -0.60$, $p < 0.001$), and Naturalness ($r(70) = -0.34$, $p < 0.01$), and a positive correlation with Arousal ($r(70) = 0.62$, $p < 0.001$). A weak negative correlation was observed for Plausibility ($r(70) = -0.25$, $p < 0.05$).

\textbf{Squeeze: }
For the \textit{Toucher}, there was a positive correlation between number of squeezes per second and Arousal ($r(70) = 0.59$, $p < 0.001$), and negative correlations with Pleasantness ($r(70) = -0.36$, $p < 0.01$) and Naturalness ($r(70) = -0.45$, $p < 0.001$). No significant correlation was observed for Valence or Plausibility ($p > 0.05$).
For the \textit{Touchee}, there was a positive correlation between number of squeezes per second and Arousal ($r(70) = 0.50$, $p < 0.001$), and a negative correlation with Pleasantness ($r(70) = -0.36$, $p < 0.01$). No significant correlation was observed for Valence, Plausibility, or Naturalness ($p > 0.05$).

\subsubsection{Subjective Measures After Experiment} 
Out of the 12 participants, only 3 reported experiencing moderate headache symptoms, while 2 out of those 3 participants reported moderate eyestrain symptoms based on the SSQ analysis. For the remaining participants, no changes were observed in SSQ scores between the pre- and post-experiment assessments.

We conducted a paired-samples t-test to compare overall presence between the User modes. There was a significant difference in presence scores between the Touchee ($M = 6.04, SD= 0.31$) and Toucher ($M = 4.96, SD = 0.7$), $t(11) = -6.02, p < 0.001, d = 0.71$.

\subsection{Experiment 2 -- Free Interaction}

No order effects were detected within participants, indicating that the sequence of the feedback conditions (V and VH) presented did not influence the results.

\begin{figure}[t!]
    \centering
    \includegraphics[width=\columnwidth]{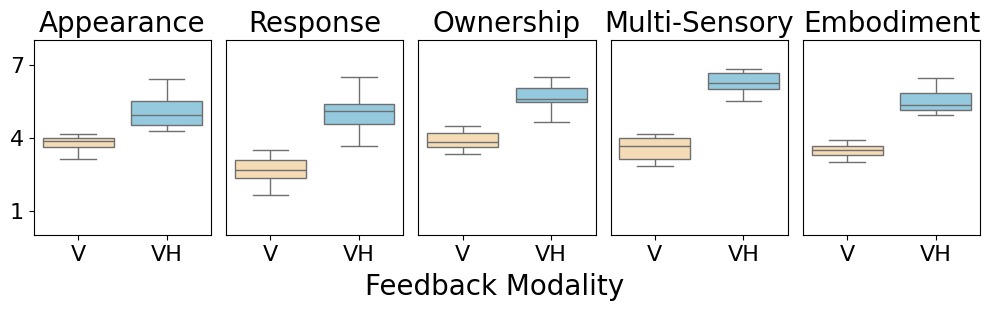}
    \caption{Embodiment scores of 12 participants across the two feedback modality conditions after the free interaction experiment}
    \label{embodiment}
\end{figure}

A paired-samples t-test revealed significant differences between the feedback conditions for all categories of the Embodiment questionnaire: Appearance ($t(11) = -6.07, p < 0.001, d = 1.75$), Response ($t(11) = -11.73, p < 0.001, d = 3.39$), Ownership ($t(11) = -10.15, p < 0.001, d = 2.93$), Multi-Sensory ($t(11) = -13.18, p < 0.001, d = 3.81$), and Embodiment ($t(11) = -11.43, p < 0.001, d = 3.30$). Scores were consistently higher for VH compared to V across all measures (Figure~\ref{embodiment}).

Paired-samples t-test conducted to compare overall presence between the Feedback conditions revealed a significant difference between V ($M = 4.09, SD= 0.78$) and VH ($M = 5.93, SD = 0.64$), $t(11) = -9.77, p < 0.001, d = 1.15$.

For System Usability, there was no significant difference ($p > 0.05$) between V ($M = 86.04, SD = 9.85$) and VH ($M = 89.58, SD = 6.64$) conditions.

None of the participants reported any changes in SSQ symptoms before and after the experiment.

The results from the HX questionnaire for the VH condition only are listed in Table~\ref{Table1}. Scores are out of a maximum of 5 points.
\begin{table}[b!]
\caption{Results from Factors of HX Questionnaire~\cite{anwar2023factors}}\label{Table1}
\begin{tabularx}{\columnwidth}{|X|c|c|}
\hline
\textbf{Question: The haptic feedback …}             & \textbf{Mean (SD)} & \textbf{Factors}                                       \\ \hline
\rowcolor[HTML]{FFFFFF} 
Felt realistic (+)                                     & 3.67 (0.49)        & \cellcolor[HTML]{FFE5E3}                               \\ \cline{1-2}
\rowcolor[HTML]{FFFFFF} 
Was believable (+)                                    & 4.33 (0.65)        & \cellcolor[HTML]{FFE5E3}                               \\ \cline{1-2}
\rowcolor[HTML]{FFFFFF} 
Was convincing (+)                                       & 4.58 (0.51)        & \multirow{-3}{*}{\cellcolor[HTML]{FFE5E3}Realism}      \\ \hline
\rowcolor[HTML]{FFFFFF} 
Felt disconnected from the rest of the experience (-)    & 1.25 (0.45)        & \cellcolor[HTML]{DAFFDA}                               \\ \cline{1-2}
\rowcolor[HTML]{FFFFFF} 
Felt out of place (-)                                    & 1.00 (0.00)        & \cellcolor[HTML]{DAFFDA}                               \\ \cline{1-2}
\rowcolor[HTML]{FFFFFF} 
Distracted me from the task (-)                          & 1.17 (0.58)        & \multirow{-3}{*}{\cellcolor[HTML]{DAFFDA}Harmony}      \\ \hline
\rowcolor[HTML]{FFFFFF} 
Enjoyable as part of the experience (+)                  & 4.67 (0.49)        & \cellcolor[HTML]{DCDEFF}                               \\ \cline{1-2}
\rowcolor[HTML]{FFFFFF} 
Felt engaging with the system (+)                        & 4.41 (0.51)        & \multirow{-2}{*}{\cellcolor[HTML]{DCDEFF}Involvement}  \\ \hline
\rowcolor[HTML]{FFFFFF} 
All felt the same (-)                                   & 1.25 (0.45)        & \cellcolor[HTML]{D8FFFE}                               \\ \cline{1-2}
\rowcolor[HTML]{FFFFFF} 
Changes depending on how things change in the system (+) & 4.50 (0.80)        & \cellcolor[HTML]{D8FFFE}                               \\ \cline{1-2}
\rowcolor[HTML]{FFFFFF} 
Reflects varying inputs and events (+)                   & 4.92 (0.29)        & \multirow{-3}{*}{\cellcolor[HTML]{D8FFFE}Expressivity} \\ \hline
\end{tabularx}
\vspace{-10pt}
\end{table}

\begin{figure}[t!]
    \centering
    \begin{subfigure}[t]{0.316\columnwidth}
        \centering
        \includegraphics[width=\columnwidth]{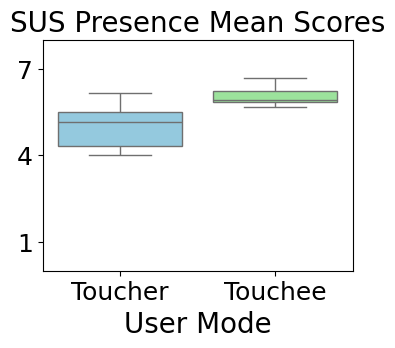}
        \caption{}
        \label{fig:prescribed}
    \end{subfigure}
    \begin{subfigure}[t]{0.316\columnwidth}
        \centering
        \includegraphics[width=\columnwidth]{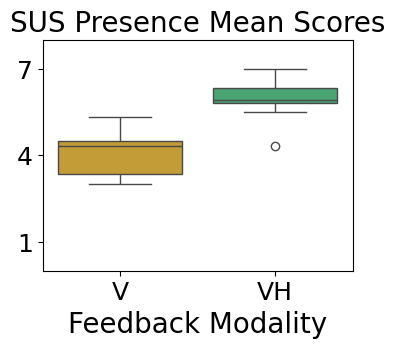}
        \caption{}
        \label{fig:freeexploration}
    \end{subfigure}
    \begin{subfigure}[t]{0.316\columnwidth}
        \centering
        \includegraphics[width=\columnwidth]{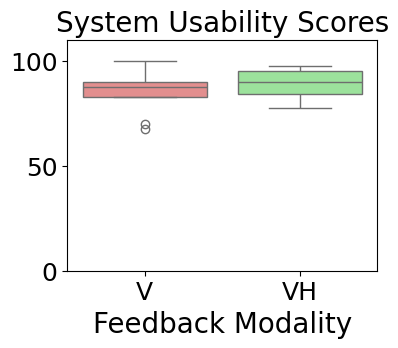}
        \caption{}
        \label{fig:usability}
    \end{subfigure}
    \caption{SUS Presence scores for (a) Prescribed Touch and (b) Free Interaction; (c) Overall VR System Usability scores across 12 participants}
    \label{fig:embodiment_combined}
\end{figure}

\section{Discussion}

No participants reported challenges or discomfort wearing the vibrotactile glove and sleeve. They fit all users, effectively accommodating large variations in hand and arm sizes.

\subsection{Experiment 1 -- Prescribed Touch}

The results across all gestures reveal the critical roles of speed at which the gesture was performed, feedback modality, and user roles in shaping emotional and sensory experiences. Speed had a consistent influence, with faster interactions heightening arousal but often reducing valence, pleasantness, and naturalness. Slower or more deliberate gestures generally produced more positive and natural experiences, fostering higher valence and pleasantness with moderate arousal levels. Normal speeds often achieved the best balance, suggesting that deviations from natural pacing -- whether faster or slower -- tend to diminish the quality of the interaction.

Multimodal feedback emerged as a key factor in enhancing emotional and sensory outcomes. Compared to visual-only feedback, visual-haptic consistently improved valence, arousal, plausibility, pleasantness, and naturalness. This effect was particularly pronounced during faster gestures, where the added plausibility, pleasantness, and naturalness  mitigated some of the negative effects of rapid interactions. For slower gestures, visual-haptic feedback further amplified the already positive experiences, making it a critical design element for fostering pleasantness and realism.

User roles also shaped the interaction dynamics, though their effects were often secondary to speed and feedback modality. Touchers and Touchees displayed distinct patterns of emotional and sensory responses. For Touchers, faster gestures often correlated with increased arousal and emotional engagement, suggesting that actively delivering rapid interactions can enhance feelings of control and involvement. However, for Touchees, faster gestures typically resulted in heightened arousal coupled with reduced valence, pleasantness, and naturalness, indicating that receiving rapid interactions may become overwhelming or uncomfortable. These differences highlight the sensory-emotional trade-offs that are particularly pronounced for Touchees and emphasize the need to account for user roles when designing interaction dynamics.

For stroking gestures, rapid motions stimulated heightened arousal but reduced emotional positivity, and naturalness, whereas slower or natural speeds elicited high valence, moderate arousal, and enhanced plausibility, pleasantness and naturalness. Visual-haptic feedback enhanced both sensory and emotional experiences. Both Touchers and Touchees displayed similar response patterns, underscoring the centrality of speed and feedback mode in shaping the experience. These findings align with the well-established role of moderate stroking speeds (1–10 $cm/s$)~\cite{strokingVelocity} in optimally engaging CT afferents, maximizing the perceived pleasantness.

Faster pokes increased arousal but decreased valence, pleasantness, and naturalness, with both slow and fast pokes seen as less natural than normal-speed ones. Touchers found rapid pokes more engaging and positive, while Touchees experienced heightened arousal and reduced pleasantness, reflecting the sensory-emotional tension of receiving fast gestures. Lack of significant correlations with plausibility for Touchers suggests speed mainly affects emotional activation, while Touchees experience both sensory and emotional impacts.

Speed similarly affected patting, with faster pats lowering valence, pleasantness, naturalness, and increasing arousal. Visual-haptic feedback mitigated these effects, especially at higher speeds, by improving positivity and comfort. Touchers reported slightly higher valence but lower arousal than Touchees, who were more sensitive to speed changes, showing greater declines in valence and pleasantness with faster pats. Correlation analyses confirmed stronger negative correlations between speed and perceived pleasantness or naturalness for Touchees, highlighting the sensory-emotional trade-offs of faster interactions.

Faster squeezing increased arousal but reduced plausibility, pleasantness, naturalness, and slightly valence. Visual-haptic feedback improved overall sensory-emotional experience, noticeably at moderate speeds. User roles had minimal direct effects, but correlations showed that for Touchers, faster squeezes increased arousal reducing naturalness, while for Touchees, there was a trade-off between increased arousal and reduced pleasantness.

Across all gestures, the interplay of speed, feedback, and user roles followed clear patterns. Faster speeds heightened arousal but at the expense of pleasantness, naturalness, and emotional positivity, while visual-haptic feedback improved plausibility and mitigated these effects to some extent. Touchers subtly derived greater emotional engagement, whereas Touchees were more sensitive to rapid gestures.
In summary, slower, deliberate gestures with multimodal feedback produced the most positive and natural experiences, particularly for Touchees. Faster gestures increase arousal and engagement but should be used carefully to avoid reducing the sensory-emotional quality. By considering the interplay of speed, feedback, and user roles, designers can optimize VR touch interactions for greater emotional engagement and user satisfaction.


The higher presence scores observed for the Touchees compared to the Touchers are likely due to the experimental instructions given to the Touchers, requiring them to perform specific gestures at different speeds. These instructions may have disrupted their sense of ``being" or immersiveness within the VR environment.

\subsection{Experiment 2 -- Free Interaction}

Visual plus haptic feedback together consistently outperformed visual-only feedback in Appearance, Response, Ownership, Multi-Sensory integration, and overall Embodiment scores. The substantial effect sizes further emphasize the magnitude of these improvements. This highlights the importance of incorporating richer haptic feedback to improve the experience of a multi-user VR system. The condition with haptic feedback included likely provides more immersive and cohesive sensory cues, which contribute to a stronger perception of body ownership and response to the interaction. Questions addressing the perception of touch sensations, interactions between the avatar and other objects or avatars in VR, and the ability to influence the user’s own body through the avatar received significantly higher ratings with haptic feedback present, compared to the visual-only condition.

The significantly higher sense of presence reported in the added haptic feedback condition 
compared to the visual-only condition 
suggests that the addition of haptic feedback enhances immersion and the feeling of ``being" in the virtual environment. This aligns with the SUS presence model, which emphasizes the role of multi-sensory inputs in strengthening spatial presence. Some participants reported that the haptic feedback made the VR environment feel more tangible and interactive, leading to a deeper sense of ``being" in the VR.

Our VR system received high usability scores, with no significant differences between using haptic feedback or not. This indicates the system is intuitive, user-friendly, and accessible, allowing easy learning and navigation. Participants efficiently used the armband, sleeves, gloves, and headset without assistance, even though 10 out of 12 had no prior VR or haptic device experience. These results suggest the system holds promise for haptic and VR research, providing a strong foundation for exploring mediated social touch.

High ratings for \textit{Harmony},
\textit{Involvement}, 
and \textit{Expressivity}, of the system's haptic feedback 
indicated that it was well-integrated, engaging, and dynamic based on interaction types. However, \textit{Realism} scores were comparatively lower,
suggesting that simple vibrations were less effective at conveying a realistic sensation. Participants noted that, while the vibrations as feedback were convincing, they did not always feel natural or fully aligned with their expectations for certain interactions.

\subsection{User Feedback}
Participants were highly enthusiastic about the VR experience, especially interacting with friends in a shared virtual space and seeing their body movements reflected in their avatars. They enjoyed interacting with multiple virtual objects and receiving touch feedback, praising the adaptive vibration feedback and system synchronization. Many highlighted the nearly imperceptible latency and the instantaneous haptic feedback. Participants particularly enjoyed the Free Interaction experiment, engaging in activities like dancing and playing games, which added fun and spontaneity to the experience.

For the Prescribed Touch experiment, among touchees, the stroking gesture on the forearm was highly praised, with many describing it as pleasant and relaxing. Participants also appreciated the haptic experience of other gestures, particularly when performed at slow or medium speeds. The ``normal" speed of gestures was often described as the most natural. Touchers similarly reported a positive experience with stroking, patting, and squeezing gestures. However, the poking gesture received less favorable feedback. We suspect this was largely due to hand tracking inaccuracies caused by the Meta Quest's limitations in detecting hand poses when participants wore the gloves.

To the open-ended question for the Touchees after the Prescribed Touch Experiment, they related the touch gestures performed as ``grabbing my arm", ``patting", ``tapping", ``poking", ``trying to grab my attention", ``caressing", ``hold and squeezing".
Some participants suggested that adding pressure cues to the hands (for touchers) and the forearm (for touchees) during squeezing and poking gestures could enhance the realism of these interactions. Notably, the patting gesture, when performed at a fast pace, was universally disliked. Participants described it as feeling unnatural, overly abrupt, or disconcerting, highlighting the need for careful consideration of intentions in mediated social touch system design.

\section{Conclusion, limitations and Future Work}

This paper introduces a standalone, device-agnostic multiplayer VR system for real-time mediated social touch with haptic feedback, designed for Meta Quest headsets and supporting up to 16 physically distant users. It integrates wireless vibrotactile feedback via gloves and sleeves with 52 ERM actuators, enabling users to perform and receive remote touch gestures. Our evaluation with 12 participants revealed the impact of interaction speed, feedback modality, and user roles on emotional and sensory experiences, alongside high system usability, presence, and embodiment scores.


This work has several limitations that highlight opportunities for future research and development. The use of ERMs in our device restricts haptic fidelity by providing only vibrotactile feedback, which cannot fully substitute for the broad range of feedback modalities. 
Moreover, our current design is limited to actuation on the hands and forearms. Expanding to full-body haptic feedback would pose significant challenges in terms of hardware design and power consumption. To address these limitations, future research must explore alternative actuation techniques capable of replicating complex sensations, such as pressure, temperature, and skin stretch, in real-time VR interactions with minimal latency.

However, our VR system is designed to be device-agnostic, enabling researchers to integrate and test their own devices within our framework. As future work, we plan to evaluate our VR system using haptic feedback provided to both the toucher and touchee from commercially available devices (e.g. Ultraleap module or bHaptic gloves), to further strengthen our claim and broaden its applicability. Furthermore, this study did not examine the influence of gender, sexual orientation, avatar characteristics, such as appearance, skin tone, and facial expressions, on user experience, an aspect worth exploring in future research. Our participant sample consisted of pairs of individuals who were familiar with each other prior to the study. Although our participant population was gender-diverse, the group size was too small to study either the effect of gender or relationship in mediated social touch. 
These effects should be studied in the future with more diverse and larger participant groups including strangers to improve the reliability and generalizability of the findings.
A key goal for future work is to make this VR system open source, enabling researchers to develop, evaluate, and compare their own haptic devices within our VR platform. The scope of this VR framework can be expanded, exploring its potential as a benchmark application to evaluate and compare tactile devices in diverse contexts~\cite{VRsuite}, beyond social touch interactions alone.

\section*{Acknowledgment}

This research was supported by the National Science Foundation under Grant No. 2047867.

\bibliographystyle{ieeetr}
\bibliography{References_Social_Touch.bib}

\begin{thebibliography}{100}

\bibitem{DH3}
G.-Z. Yang, ``How could robotics help establish a new norm after covid-19?,'' 2021.

\bibitem{DH4}
D.~Prattichizzo, ``Beyond the pandemic: The role of haptics in defining the new normal,'' {\em IEEE Transactions on Haptics}, vol.~14, no.~1, pp.~1--1, 2021.

\bibitem{DH5}
C.~S. De~Figueiredo, P.~C. Sandre, L.~C.~L. Portugal, T.~M{\'a}zala-de Oliveira, L.~da~Silva~Chagas, {\'I}.~Raony, E.~S. Ferreira, E.~Giestal-de Araujo, A.~A. Dos~Santos, and P.~O.-S. Bomfim, ``Covid-19 pandemic impact on children and adolescents' mental health: Biological, environmental, and social factors,'' {\em Progress in Neuro-Psychopharmacology and Biological Psychiatry}, vol.~106, p.~110171, 2021.

\bibitem{DH6}
K.~Usher, J.~Durkin, and N.~Bhullar, ``The covid-19 pandemic and mental health impacts,'' {\em International journal of mental health nursing}, vol.~29, no.~3, p.~315, 2020.

\bibitem{DH7}
M.~Wang, ``Social {VR}: a new form of social communication in the future or a beautiful illusion?,'' in {\em Journal of Physics: Conference Series}, vol.~1518, p.~012032, IOP Publishing, 2020.

\bibitem{DH8}
A.~B{\"o}nsch, S.~Radke, H.~Overath, L.~M. Asch{\'e}, J.~Wendt, T.~Vierjahn, U.~Habel, and T.~W. Kuhlen, ``Social {VR}: How personal space is affected by virtual agents' emotions,'' in {\em 2018 IEEE conference on virtual reality and 3D user interfaces (VR)}, pp.~199--206, IEEE, 2018.

\bibitem{DH9}
J.~Li, V.~Vinayagamoorthy, J.~Williamson, D.~A. Shamma, and P.~Cesar, ``Social {VR}: A new medium for remote communication and collaboration,'' in {\em Extended Abstracts of the 2021 CHI Conference on Human Factors in Computing Systems}, pp.~1--6, 2021.

\bibitem{qiu2023vigather}
H.~Qiu, P.~Streli, T.~Luong, C.~Gebhardt, and C.~Holz, ``Vigather: Inclusive virtual conferencing with a joint experience across traditional screen devices and mixed reality headsets,'' {\em Proceedings of the ACM on Human-Computer Interaction}, vol.~7, no.~MHCI, pp.~1--27, 2023.

\bibitem{DH10}
M.~J. Hertenstein, J.~M. Verkamp, A.~M. Kerestes, and R.~M. Holmes, ``The communicative functions of touch in humans, nonhuman primates, and rats: a review and synthesis of the empirical research,'' {\em Genetic, social, and general psychology monographs}, vol.~132, no.~1, pp.~5--94, 2006.

\bibitem{DH11}
S.~Yohanan and K.~E. MacLean, ``The role of affective touch in human-robot interaction: Human intent and expectations in touching the haptic creature,'' {\em International Journal of Social Robotics}, vol.~4, pp.~163--180, 2012.

\bibitem{DH12}
M.~Argyle, {\em Bodily communication}.
\newblock Routledge, 2013.

\bibitem{DH13}
M.-J. Han, C.-H. Lin, and K.-T. Song, ``Robotic emotional expression generation based on mood transition and personality model,'' {\em IEEE transactions on cybernetics}, vol.~43, no.~4, pp.~1290--1303, 2012.

\bibitem{DH14}
R.~Stock-Homburg, {\em Negative interaction spirals during service encounters: insights from human-human and human-robot interactions}.
\newblock PhD thesis, FernUniversit{\"a}t Hagen, 2018.

\bibitem{ST8}
C.~J. Cascio, D.~Moore, and F.~McGlone, ``Social touch and human development,'' {\em Developmental cognitive neuroscience}, vol.~35, pp.~5--11, 2019.

\bibitem{ST27}
G.~Huisman, ``Social touch technology: A survey of haptic technology for social touch,'' {\em IEEE transactions on haptics}, vol.~10, no.~3, pp.~391--408, 2017.

\bibitem{wei2023mediated}
Q.~Wei, M.~Li, and J.~Hu, ``Mediated social touch with mobile devices: A review of designs and evaluations,'' {\em IEEE Transactions on Haptics}, 2023.

\bibitem{DH43}
Y.~K. Dwivedi, L.~Hughes, A.~M. Baabdullah, S.~Ribeiro-Navarrete, M.~Giannakis, M.~M. Al-Debei, D.~Dennehy, B.~Metri, D.~Buhalis, C.~M. Cheung, {\em et~al.}, ``Metaverse beyond the hype: Multidisciplinary perspectives on emerging challenges, opportunities, and agenda for research, practice and policy,'' {\em International journal of information management}, vol.~66, p.~102542, 2022.

\bibitem{DH44}
J.~D.~N. Dionisio, W.~G.~B. Iii, and R.~Gilbert, ``3d virtual worlds and the metaverse: Current status and future possibilities,'' {\em ACM computing surveys (CSUR)}, vol.~45, no.~3, pp.~1--38, 2013.

\bibitem{ALT17}
F.~Arafsha, K.~M. Alam, and A.~El~Saddik, ``Design and development of a user centric affective haptic jacket,'' {\em Multimedia Tools and Applications}, vol.~74, pp.~3035--3052, 2015.

\bibitem{ALT18}
Akshita, H.~Alagarai~Sampath, B.~Indurkhya, E.~Lee, and Y.~Bae, ``Towards multimodal affective feedback: Interaction between visual and haptic modalities,'' in {\em Proceedings of the 33rd Annual ACM Conference on Human Factors in Computing Systems}, pp.~2043--2052, 2015.

\bibitem{ALT19}
J.~Mullenbach, C.~Shultz, J.~E. Colgate, and A.~M. Piper, ``Exploring affective communication through variable-friction surface haptics,'' in {\em Proceedings of the SIGCHI Conference on Human Factors in Computing Systems}, pp.~3963--3972, 2014.

\bibitem{ALT20}
Y.~Yoo, T.~Yoo, J.~Kong, and S.~Choi, ``Emotional responses of tactile icons: Effects of amplitude, frequency, duration, and envelope,'' in {\em 2015 IEEE World Haptics Conference (WHC)}, pp.~235--240, IEEE, 2015.

\bibitem{ALT21}
Y.~Chandra, R.~Peiris, and K.~Minamizawa, ``Affective haptic furniture: Directional vibration pattern to regulate emotion,'' in {\em Proceedings of the 2018 ACM International Joint Conference and 2018 International Symposium on Pervasive and Ubiquitous Computing and Wearable Computers}, pp.~25--28, 2018.

\bibitem{ALT22}
A.~Israr, S.~Zhao, K.~Schwalje, R.~Klatzky, and J.~Lehman, ``Feel effects: enriching storytelling with haptic feedback,'' {\em ACM Transactions on Applied Perception (TAP)}, vol.~11, no.~3, pp.~1--17, 2014.

\bibitem{ALT27}
C.~Rognon, B.~Stephens-Fripp, J.~Hartcher-O'Brien, B.~Rost, and A.~Israr, ``Linking haptic parameters to the emotional space for mediated social touch,'' {\em Frontiers in Computer Science}, vol.~4, p.~826545, 2022.

\bibitem{suvilehto2015topography}
J.~T. Suvilehto, E.~Glerean, R.~I. Dunbar, R.~Hari, and L.~Nummenmaa, ``Topography of social touching depends on emotional bonds between humans,'' {\em Proceedings of the National Academy of Sciences}, vol.~112, no.~45, pp.~13811--13816, 2015.

\bibitem{ALT16}
G.~Huisman, A.~D. Frederiks, J.~B. Van~Erp, and D.~K. Heylen, ``Simulating affective touch: Using a vibrotactile array to generate pleasant stroking sensations,'' in {\em Haptics: Perception, Devices, Control, and Applications: 10th International Conference, EuroHaptics 2016, London, UK, July 4-7, 2016, Proceedings, Part II 10}, pp.~240--250, Springer, 2016.

\bibitem{ALT28}
G.~Huisman, A.~D. Frederiks, B.~Van~Dijk, D.~Hevlen, and B.~Kr{\"o}se, ``The tasst: Tactile sleeve for social touch,'' in {\em 2013 World Haptics Conference (WHC)}, pp.~211--216, IEEE, 2013.

\bibitem{ALT29}
J.-J. Cabibihan, L.~Zheng, and C.~K.~T. Cher, ``Affective tele-touch,'' in {\em Social Robotics: 4th International Conference, ICSR 2012, Chengdu, China, October 29-31, 2012. Proceedings 4}, pp.~348--356, Springer, 2012.

\bibitem{ALT30}
X.~L. Cang, A.~Israr, and K.~E. MacLean, ``When is a haptic message like an inside joke? digitally mediated emotive communication builds on shared history,'' {\em IEEE Transactions on Affective Computing}, vol.~14, no.~1, pp.~732--746, 2023.

\bibitem{kirchner2023phantom}
R.~Kirchner, R.~Rosenkranz, B.~G. Sousa, S.-C. Li, and M.~E. Altinsoy, ``Phantom illusion based vibrotactile rendering of affective touch patterns,'' {\em IEEE Transactions on Haptics}, 2023.

\bibitem{ALT31}
S.~Muthukumarana, D.~S. Elvitigala, J.~P. Forero~Cortes, D.~J. Matthies, and S.~Nanayakkara, ``Touch me gently: recreating the perception of touch using a shape-memory alloy matrix,'' in {\em Proceedings of the 2020 CHI Conference on Human Factors in Computing Systems}, pp.~1--12, 2020.

\bibitem{wang2012keep}
R.~Wang, F.~Quek, D.~Tatar, K.~S. Teh, and A.~Cheok, ``Keep in touch: channel, expectation and experience,'' in {\em Proceedings of the SIGCHI Conference on Human Factors in Computing Systems}, pp.~139--148, 2012.

\bibitem{he2024affective}
S.~He, H.~Zeng, M.~Xue, G.~Huang, C.~Yao, and F.~Ying, ``Affective stroking: Design thermal mid-air tactile for assisting people in stress regulation,'' {\em Applied Sciences}, vol.~14, no.~20, p.~9494, 2024.

\bibitem{ALT32}
N.~Ferguson, M.~E. Cansev, A.~Dwivedi, and P.~Beckerle, ``Design of a wearable haptic device to mediate affective touch with a matrix of linear actuators,'' in {\em International Conference on System-Integrated Intelligence}, pp.~507--517, Springer, 2022.

\bibitem{ALT33}
W.~Wu and H.~Culbertson, ``Wearable haptic pneumatic device for creating the illusion of lateral motion on the arm,'' in {\em 2019 IEEE World Haptics Conference (WHC)}, pp.~193--198, IEEE, 2019.

\bibitem{ALT7}
X.~Zhu, T.~Feng, and H.~Culbertson, ``Understanding the effect of speed on human emotion perception in mediated social touch using voice coil actuators,'' {\em Frontiers in Computer Science}, vol.~4, p.~826637, 2022.

\bibitem{ALT34}
H.~Culbertson, C.~M. Nunez, A.~Israr, F.~Lau, F.~Abnousi, and A.~M. Okamura, ``A social haptic device to create continuous lateral motion using sequential normal indentation,'' in {\em 2018 IEEE Haptics Symposium (HAPTICS)}, pp.~32--39, IEEE, 2018.

\bibitem{ALT45}
S.~J. Lederman and L.~A. Jones, ``Tactile and haptic illusions,'' {\em IEEE Transactions on Haptics}, vol.~4, no.~4, pp.~273--294, 2011.

\bibitem{ALT44}
O.~S. Schneider, A.~Israr, and K.~E. MacLean, ``Tactile animation by direct manipulation of grid displays,'' in {\em Proceedings of the 28th annual ACM symposium on user interface software \& technology}, pp.~21--30, 2015.

\bibitem{tactilebrush}
A.~Israr and I.~Poupyrev, ``Tactile brush: drawing on skin with a tactile grid display,'' in {\em Proceedings of the SIGCHI Conference on Human Factors in Computing Systems}, pp.~2019--2028, 2011.

\bibitem{rantala2013touch}
J.~Rantala, K.~Salminen, R.~Raisamo, and V.~Surakka, ``Touch gestures in communicating emotional intention via vibrotactile stimulation,'' {\em International Journal of Human-Computer Studies}, vol.~71, no.~6, pp.~679--690, 2013.

\bibitem{bailenson2007virtual}
J.~N. Bailenson, N.~Yee, S.~Brave, D.~Merget, and D.~Koslow, ``Virtual interpersonal touch: Expressing and recognizing emotions through haptic devices,'' {\em Human--Computer Interaction}, vol.~22, no.~3, pp.~325--353, 2007.

\bibitem{ALT35}
M.~Salvato, S.~R. Williams, C.~M. Nunez, X.~Zhu, A.~Israr, F.~Lau, K.~Klumb, F.~Abnousi, A.~M. Okamura, and H.~Culbertson, ``Data-driven sparse skin stimulation can convey social touch information to humans,'' {\em IEEE Transactions on Haptics}, vol.~15, no.~2, pp.~392--404, 2021.

\bibitem{huisman2013towards}
G.~Huisman and A.~Darriba~Frederiks, ``Towards tactile expressions of emotion through mediated touch,'' in {\em CHI'13 Extended Abstracts on Human Factors in Computing Systems}, pp.~1575--1580, 2013.

\bibitem{smith2007communicating}
J.~Smith and K.~MacLean, ``Communicating emotion through a haptic link: Design space and methodology,'' {\em International Journal of Human-Computer Studies}, vol.~65, no.~4, pp.~376--387, 2007.

\bibitem{bailenson2005digital}
J.~N. Bailenson and N.~Yee, ``Digital chameleons: Automatic assimilation of nonverbal gestures in immersive virtual environments,'' {\em Psychological science}, vol.~16, no.~10, pp.~814--819, 2005.

\bibitem{maloney2020talking}
D.~Maloney, G.~Freeman, and D.~Y. Wohn, ``" talking without a voice" understanding non-verbal communication in social virtual reality,'' {\em Proceedings of the ACM on Human-Computer Interaction}, vol.~4, no.~CSCW2, pp.~1--25, 2020.

\bibitem{zamuner2013role}
E.~Zamuner, ``The role of the visual system in emotion perception,'' {\em Acta Analytica}, vol.~28, pp.~179--187, 2013.

\bibitem{blom2021perceiving}
S.~S.~A. Blom, H.~Aarts, and G.~R. Semin, ``Perceiving emotions in visual stimuli: social verbal context facilitates emotion detection of words but not of faces,'' {\em Experimental Brain Research}, vol.~239, pp.~413--423, 2021.

\bibitem{ST28}
W.~A. IJsselsteijn, Y.~A.~W. de~Kort, and A.~Haans, ``Is this my hand i see before me? the rubber hand illusion in reality, virtual reality, and mixed reality,'' {\em Presence: Teleoperators and Virtual Environments}, vol.~15, no.~4, pp.~455--464, 2006.

\bibitem{ST34}
K.~Kilteni, R.~Groten, and M.~Slater, ``The sense of embodiment in virtual reality,'' {\em Presence: Teleoperators and Virtual Environments}, vol.~21, no.~4, pp.~373--387, 2012.

\bibitem{ST40}
A.~Maselli and M.~Slater, ``The building blocks of the full body ownership illusion,'' {\em Frontiers in human neuroscience}, vol.~7, p.~83, 2013.

\bibitem{ST43}
V.~I. Petkova and H.~H. Ehrsson, ``If i were you: perceptual illusion of body swapping,'' {\em PloS one}, vol.~3, no.~12, p.~e3832, 2008.

\bibitem{ST49}
M.~Slater, D.~P{\'e}rez~Marcos, H.~Ehrsson, and M.~V. Sanchez-Vives, ``Towards a digital body: the virtual arm illusion,'' {\em Frontiers in human neuroscience}, vol.~2, p.~181, 2008.

\bibitem{ST50}
M.~Slater, B.~Spanlang, M.~V. Sanchez-Vives, and O.~Blanke, ``First person experience of body transfer in virtual reality,'' {\em PloS one}, vol.~5, no.~5, p.~e10564, 2010.

\bibitem{jacucci2024haptics}
G.~Jacucci, A.~Bellucci, I.~Ahmed, V.~Harjunen, M.~Spape, and N.~Ravaja, ``Haptics in social interaction with agents and avatars in virtual reality: a systematic review,'' {\em Virtual Reality}, vol.~28, no.~4, p.~170, 2024.

\bibitem{genay2021being}
A.~Genay, A.~L{\'e}cuyer, and M.~Hachet, ``Being an avatar “for real”: a survey on virtual embodiment in augmented reality,'' {\em IEEE Transactions on Visualization and Computer Graphics}, vol.~28, no.~12, pp.~5071--5090, 2021.

\bibitem{ST18}
A.~Haans and W.~IJsselsteijn, ``Mediated social touch: a review of current research and future directions,'' {\em Virtual Reality}, vol.~9, pp.~149--159, 2006.

\bibitem{ST60}
J.~B. Van~Erp and A.~Toet, ``Social touch in human--computer interaction,'' {\em Frontiers in digital humanities}, vol.~2, p.~2, 2015.

\bibitem{ST26}
M.~Hoppe, B.~Rossmy, D.~P. Neumann, S.~Streuber, A.~Schmidt, and T.-K. Machulla, ``A human touch: Social touch increases the perceived human-likeness of agents in virtual reality,'' in {\em Proceedings of the 2020 CHI conference on human factors in computing systems}, pp.~1--11, 2020.

\bibitem{ST2}
J.~N. Bailenson and N.~Yee, ``Virtual interpersonal touch: Haptic interaction and copresence in collaborative virtual environments,'' {\em Multimedia Tools and Applications}, vol.~37, pp.~5--14, 2008.

\bibitem{ST59}
L.~Tremblay, M.~Roy-Vaillancourt, B.~Chebbi, S.~Bouchard, M.~Daoust, J.~D{\'e}nomm{\'e}e, and M.~Thorpe, ``Body image and anti-fat attitudes: an experimental study using a haptic virtual reality environment to replicate human touch,'' {\em Cyberpsychology, behavior, and social networking}, vol.~19, no.~2, pp.~100--106, 2016.

\bibitem{ST15}
A.~Gallace and C.~Spence, ``The science of interpersonal touch: an overview,'' {\em Neuroscience \& Biobehavioral Reviews}, vol.~34, no.~2, pp.~246--259, 2010.

\bibitem{ST46}
V.~Russo, C.~Ottaviani, and G.~F. Spitoni, ``Affective touch: A meta-analysis on sex differences,'' {\em Neuroscience \& Biobehavioral Reviews}, vol.~108, pp.~445--452, 2020.

\bibitem{ST14}
M.~Fusaro, M.~P. Lisi, G.~Tieri, and S.~M. Aglioti, ``Touched by vision: how heterosexual, gay, and lesbian people react to the view of their avatar being caressed on taboo body parts,'' 2020.

\bibitem{ST55}
J.~Swidrak and G.~Pochwatko, ``Being touched by a virtual human. relationships between heart rate, gender, social status, and compliance.,'' in {\em Proceedings of the 19th ACM international conference on intelligent virtual agents}, pp.~49--55, 2019.

\bibitem{ST21}
V.~J. Harjunen, M.~Spap{\'e}, I.~Ahmed, G.~Jacucci, and N.~Ravaja, ``Persuaded by the machine: The effect of virtual nonverbal cues and individual differences on compliance in economic bargaining,'' {\em Computers in Human Behavior}, vol.~87, pp.~384--394, 2018.

\bibitem{ST65}
Y.~Zhao, J.~Ham, and J.~van~der Vlist, ``Persuasive virtual touch: The effect of artificial social touch on shopping behavior in virtual reality,'' in {\em Symbiotic Interaction: 6th International Workshop, Symbiotic 2017, Eindhoven, The Netherlands, December 18--19, 2017, Revised Selected Papers 6}, pp.~98--109, Springer, 2018.

\bibitem{ST20}
V.~J. Harjunen, M.~Spap{\'e}, I.~Ahmed, G.~Jacucci, and N.~Ravaja, ``Individual differences in affective touch: Behavioral inhibition and gender define how an interpersonal touch is perceived,'' {\em Personality and Individual Differences}, vol.~107, pp.~88--95, 2017.

\bibitem{toucherFeedback}
M.~Maunsbach, K.~Hornbæk, and H.~Seifi, ``Mediated social touching: Haptic feedback affects social experience of touch initiators,'' in {\em 2023 IEEE World Haptics Conference (WHC)}, pp.~93--100, 2023.

\bibitem{DH32}
Y.~Kunii and H.~Hashimoto, ``Tele-handshake using handshake device,'' in {\em Proceedings of IECON'95-21st Annual Conference on IEEE Industrial Electronics}, vol.~1, pp.~179--182, IEEE, 1995.

\bibitem{DH33}
H.~Hashimoto and S.~Manoratkul, ``Tele-handshake through the internet,'' in {\em Proceedings 5th IEEE International Workshop on Robot and Human Communication. RO-MAN'96 TSUKUBA}, pp.~90--95, IEEE, 1996.

\bibitem{DH34}
K.~Ouchi and S.~Hashimoto, ``Handshake telephone system to communicate with voice and force,'' in {\em Proceedings 6th IEEE International Workshop on Robot and Human Communication. RO-MAN'97 SENDAI}, pp.~466--471, IEEE, 1997.

\bibitem{DH35}
H.~Nakanishi, K.~Tanaka, and Y.~Wada, ``Remote handshaking: touch enhances video-mediated social telepresence,'' in {\em Proceedings of the SIGCHI conference on human factors in computing systems}, pp.~2143--2152, 2014.

\bibitem{DH36}
N.~Pedemonte, T.~Lalibert{\'e}, and C.~Gosselin, ``Design, control, and experimental validation of a handshaking reactive robotic interface,'' {\em Journal of Mechanisms and Robotics}, vol.~8, no.~1, p.~011020, 2016.

\bibitem{DH37}
N.~Pedemonte, T.~Lalibert{\'e}, and C.~Gosselin, ``A haptic bilateral system for the remote human--human handshake,'' {\em Journal of Dynamic Systems, Measurement, and Control}, vol.~139, no.~4, p.~044503, 2017.

\bibitem{DH38}
S.-M. Park and Y.-G. Kim, ``A metaverse: Taxonomy, components, applications, and open challenges,'' {\em IEEE access}, vol.~10, pp.~4209--4251, 2022.

\bibitem{tong2024distant}
Q.~Tong, W.~Wei, Y.~Guo, T.~Jin, Z.~Wang, H.~Zhang, Y.~Zhang, and D.~Wang, ``Distant handshakes: Conveying social intentions through multi-modal soft haptic gloves,'' {\em IEEE Transactions on Affective Computing}, 2024.

\bibitem{DH39}
M.~O. Alhalabi and S.~Horiguchi, ``Tele-handshake: a cooperative shared haptic virtual environment,'' in {\em Proceedings of Eurohaptics}, vol.~2001, pp.~60--64, Citeseer, 2001.

\bibitem{DH40}
T.~Miyoshi, Y.~Ueno, K.~Kawase, Y.~Matsuda, Y.~Ogawa, K.~Takemori, and K.~Terashima, ``Development of handshake gadget and exhibition in niconico chokaigi,'' {\em Haptic Interaction: Perception, Devices and Applications}, pp.~267--272, 2015.

\bibitem{DH42}
X.~Gu, Y.~Zhang, W.~Sun, Y.~Bian, D.~Zhou, and P.~O. Kristensson, ``Dexmo: An inexpensive and lightweight mechanical exoskeleton for motion capture and force feedback in vr,'' in {\em Proceedings of the 2016 CHI Conference on Human Factors in Computing Systems}, pp.~1991--1995, 2016.

\bibitem{sykownik2020experience}
P.~Sykownik and M.~Masuch, ``The experience of social touch in multi-user virtual reality,'' in {\em Proceedings of the 26th ACM Symposium on Virtual Reality Software and Technology}, pp.~1--11, 2020.

\bibitem{ST7}
P.~Bourdin, J.~M.~T. Sanahuja, C.~C. Moya, P.~Haggard, and M.~Slater, ``Persuading people in a remote destination to sing by beaming there,'' in {\em Proceedings of the 19th ACM Symposium on Virtual Reality Software and Technology}, pp.~123--132, 2013.

\bibitem{peng2024impact}
D.~Peng, T.~Person, X.~Shen, Y.~S. Pai, G.~Barbareschi, S.~Li, and K.~Minamizawa, ``Impact of vibrotactile triggers on mental well-being through asmr experience in vr,'' in {\em International Conference on Human Haptic Sensing and Touch Enabled Computer Applications}, pp.~398--410, Springer, 2024.

\bibitem{ariza2016inducing}
O.~Ariza, J.~Freiwald, N.~Laage, M.~Feist, M.~Salloum, G.~Bruder, and F.~Steinicke, ``Inducing body-transfer illusions in vr by providing brief phases of visual-tactile stimulation,'' in {\em Proceedings of the 2016 Symposium on Spatial User Interaction}, pp.~61--68, 2016.

\bibitem{israrForearm}
A.~Israr and F.~Abnousi, ``Towards pleasant touch: Vibrotactile grids for social touch interactions,'' in {\em Extended Abstracts of the 2018 CHI Conference on Human Factors in Computing Systems}, CHI EA '18, p.~1–6, Association for Computing Machinery, 2018.

\bibitem{giannopoulos2012touching}
E.~Giannopoulos, A.~Pomes, and M.~Slater, ``Touching the void: exploring virtual objects through a vibrotactile glove,'' {\em International Journal of Virtual Reality}, vol.~11, no.~3, pp.~19--24, 2012.

\bibitem{thompson2011effect}
E.~H. Thompson and J.~A. Hampton, ``The effect of relationship status on communicating emotions through touch,'' {\em Cognition and Emotion}, vol.~25, no.~2, pp.~295--306, 2011.

\bibitem{hertenstein2009communication}
M.~J. Hertenstein, R.~Holmes, M.~McCullough, and D.~Keltner, ``The communication of emotion via touch.,'' {\em Emotion}, vol.~9, no.~4, p.~566, 2009.

\bibitem{webb2015individual}
A.~Webb and J.~Peck, ``Individual differences in interpersonal touch: On the development, validation, and use of the “comfort with interpersonal touch”(cit) scale,'' {\em Journal of consumer psychology}, vol.~25, no.~1, pp.~60--77, 2015.

\bibitem{kennedy1993simulator}
R.~S. Kennedy, N.~E. Lane, K.~S. Berbaum, and M.~G. Lilienthal, ``Simulator sickness questionnaire: An enhanced method for quantifying simulator sickness,'' {\em The international journal of aviation psychology}, 1993.

\bibitem{toet2019emojigrid}
A.~Toet and J.~Van~Erp, ``The emojigrid as a tool to assess experienced and perceived emotions. psych 1 (1): 469--481,'' 2019.

\bibitem{slater1998}
M.~Slater, A.~Steed, J.~McCarthy, and F.~Maringelli, ``The influence of body movement on subjective presence in virtual environments,'' {\em Human factors}, vol.~40, no.~3, pp.~469--477, 1998.

\bibitem{usoh1999}
M.~Usoh, K.~Arthur, M.~C. Whitton, R.~Bastos, A.~Steed, M.~Slater, and F.~P. Brooks~Jr, ``Walking> walking-in-place> flying, in virtual environments,'' in {\em Proceedings of the 26th annual conference on Computer graphics and interactive techniques}, pp.~359--364, 1999.

\bibitem{peck2021avatar}
T.~C. Peck and M.~Gonzalez-Franco, ``Avatar embodiment. a standardized questionnaire,'' {\em Frontiers in Virtual Reality}, vol.~1, p.~575943, 2021.

\bibitem{brooke1996sus}
J.~Brooke {\em et~al.}, ``Sus-a quick and dirty usability scale,'' {\em Usability evaluation in industry}, vol.~189, no.~194, pp.~4--7, 1996.

\bibitem{anwar2023factors}
A.~Anwar, T.~Shi, and O.~Schneider, ``Factors of haptic experience across multiple haptic modalities,'' in {\em Proceedings of the 2023 CHI Conference on Human Factors in Computing Systems}, pp.~1--12, 2023.

\bibitem{strokingVelocity}
L.~S. L{\"o}ken, J.~Wessberg, F.~McGlone, and H.~Olausson, ``Coding of pleasant touch by unmyelinated afferents in humans,'' {\em Nature neuroscience}, vol.~12, no.~5, pp.~547--548, 2009.

\bibitem{VRsuite}
P.~Banerjee, E.~Muschter, H.~Singh, B.~Weber, and T.~Hulin, ``Towards a vr evaluation suite for tactile displays in telerobotic space missions,'' in {\em 2023 IEEE Aerospace Conference}, pp.~1--12, IEEE, 2023.

\end{thebibliography}

\end{document}